\documentclass[journal,10pt]{IEEEtran}               
\usepackage[T1]{fontenc}

\usepackage{amsmath}
\usepackage{amssymb}
\usepackage{stmaryrd}
\usepackage{epic,eepic}
\usepackage{theorem}
\usepackage{pifont}  
\usepackage{euscript}
\usepackage{calc}
\usepackage[titles]{tocloft}

\usepackage[usenames]{color}          
\usepackage{xcolor}
\definecolor{Brown}{rgb}{0.55,0.0,0.10}
\definecolor{dgreen}{rgb}{0.00,0.56,0.00}
\definecolor{vertmoinsfonce}{rgb}{0.00,0.50,0.00}
\definecolor{vert}{rgb}{0.00,0.60,0.00}
\definecolor{llightggray}{rgb}{0.97,0.97,0.97}
\definecolor{lightggray}{rgb}{0.9,0.9,0.9}
\definecolor{ggray}{rgb}{0.5,0.5,0.5}
\definecolor{darkggray}{rgb}{0.25,0.25,0.25}
\definecolor{ddarkggray}{rgb}{0.1,0.1,0.1}
\definecolor{bleu}{rgb}{0.00,0.00,1.00}
\definecolor{darkblue}{rgb}{0,0,0.7}

\usepackage{epsf}           
\usepackage{graphicx}       
\usepackage{epsfig}         
\usepackage{pstricks}
\usepackage{pstricks-add}
\usepackage{pst-plot}
\usepackage{dsfont}
\theoremheaderfont{\color{black}\normalfont\bfseries}

\newtheorem{lemma}{Lemma}
\newtheorem{theorem}{Theorem}[section]
\newtheorem{definition}[theorem]{Definition}

\newtheorem{proposition}[theorem]{Proposition}

\theoremstyle{plain}{\theorembodyfont{\rmfamily}%
}
\theoremstyle{plain}{\theorembodyfont{\rmfamily}%
}
\theoremstyle{plain}{
\theorembodyfont{\rmfamily}

	\newtheorem{remark}[theorem]{Remark}

	}

\newcommand{\R}{\mathbb{R}}

\newcommand{\N}{\mathbb{N}}

\newcommand{\E}{\mathbb{E}}
\newcommand{\Q}{\mathbb{Q}}

\newcommand{\QQ}{\mathcal{Q}}

\newcommand{\PP}{\mathcal{P}}

\newcommand{\mc}{\mathcal}
\newcommand{\bd}{\textbf}

{\Large\normalsize}
{\Large\normalsize}

\ifCLASSINFOpdf
\else
\fi

\begin{document}


\title{Information Design for Strategic Coordination of Autonomous Devices with Non-Aligned Utilities}

%
%
%


\author{\IEEEauthorblockN{Ma\"{e}l Le Treust\IEEEauthorrefmark{1} and 
Tristan Tomala \IEEEauthorrefmark{2}}\\
\IEEEauthorblockA{\IEEEauthorrefmark{1}
ETIS, UMR 8051 / ENSEA, Université Cergy-Pontoise, CNRS,\\
6, avenue du Ponceau, 95014 Cergy-Pontoise CEDEX, FRANCE\\
Email: mael.le-treust@ensea.fr}
\thanks{\IEEEauthorrefmark{1} Ma\"{e}l Le Treust acknowledges financial support from INS2I CNRS through projects  JCJC CoReDe 2015 and PEPS StrategicCoo 2016.}\\
\IEEEauthorblockA{\IEEEauthorrefmark{2}
HEC Paris, GREGHEC UMR 2959\\
1 rue de la Libération, 78351 Jouy-en-Josas CEDEX, FRANCE\\
Email: tomala@hec.fr}
\thanks{\IEEEauthorrefmark{2} Tristan Tomala acknowledges financial support from the HEC foundation.}}


\maketitle

%
\IEEEpeerreviewmaketitle





\begin{abstract}

In this paper, we investigate the coordination of autonomous devices with non-aligned utility functions. Both encoder and decoder are considered as players, that choose the encoding and the decoding in order to maximize their long-run utility functions. The topology of the point-to-point network under investigation, suggests that the decoder implements a strategy, knowing in advance the strategy of the encoder. We characterize the encoding and decoding functions that form an equilibrium, by using empirical coordination. The equilibrium solution is related to an auxiliary game in which both players choose some conditional distributions in order to maximize their expected utilities. This problem is closely related to the literature on ``Information Design'' in Game Theory. We also characterize the set of posterior distributions that are compatible with a rate-limited channel between the encoder and the decoder. Finally, we provide an example of non-aligned utility functions corresponding to parallel fading multiple access channels.

\end{abstract}


\section{Introduction}\label{sec:Introduction}

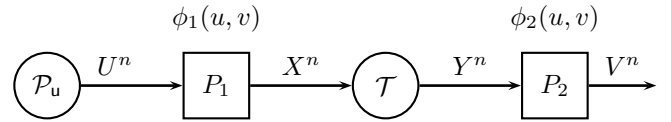
\begin{figure}[!ht]
\begin{center}
\psset{xunit=0.9cm,yunit=0.9cm}
\begin{pspicture}(0,-0.1)(8.5,1.7)
\pscircle(0,0.5){0.45}
\psframe(2,0)(3,1)
\pscircle(5,0.5){0.45}
\psframe(7,0)(8,1)
\psline[linewidth=1pt]{->}(0.5,0.5)(2,0.5)
\psline[linewidth=1pt]{->}(3,0.5)(4.5,0.5)
\psline[linewidth=1pt]{->}(5.5,0.5)(7,0.5)
\psline[linewidth=1pt]{->}(8,0.5)(9,0.5)
\rput[u](1,0.8){$U^{n}$}
\rput[u](3.75,0.8){$X^n$}
\rput[u](6.25,0.8){$Y^n$}
\rput[u](8.5,0.8){$V^n$}
\rput(0,0.5){$\PP_{\sf{u}}$}
\rput(5,0.5){$\mc{T}$}
\rput(2.5,0.5){$P_1$}
\rput(7.5,0.5){$P_2$}
\rput(2.5,1.5){$\phi_1(u,v)$}
\rput(7.5,1.5){$\phi_2(u,v)$}
\end{pspicture}
\caption{Strategic Empirical Coordination: The information source is i.i.d. $\PP_{\sf{u}}$ and the channel $\mc{T}$ is memoryless. The encoder $P_1$ and the decoder $P_2$ are players, endowed with non-aligned utility functions $\phi_1(u,v) \in \R$ and $\phi_2(u,v) \in \R $, depending on the source $U$ and decoder's action $V$. }
\label{fig:StrategicEmpiricalCoordination}
\end{center}
\end{figure}

In this paper, we investigate the coordination of autonomous devices with non-aligned utility functions. We consider a  point-to-point network, depicted in Fig. \ref{fig:StrategicEmpiricalCoordination}, with an i.i.d. information source $\PP_{\sf{u}}(u)$, an encoder $P_1$, a memoryless channel $\mc{T}(y|x)$ and a decoder $P_2$. The encoder and the decoder are considered as players, endowed with utility functions $\phi_1(u,v) \in \R$ and $\phi_2(u,v) \in \R $. Both utilities depend on the source symbol $u\in\mc{U}$ and on the action $v\in\mc{V}$ of player $P_2$, the decoder. In the $n$-stage game, the players choose the optimal encoding and the decoding functions. The accumulated utilities are characterized by using the empirical coordination of the random variables $(U,V)$. 

The problem of empirical coordination was investigated in both literatures of Game Theory \cite{GoHerNey06}, \cite{GossnerTomala07}, \cite{GossnerTomala06}, \cite{GossnerLarakiTomala09}, \cite{GossnerVieille02} and Information Theory \cite{KramerSavari07}, \cite{CuffPermuterCover10}, \cite{Cuff(ImplicitCoordination)11}, 
\cite{CuffSchieler11}, \cite{LetreustZaidiLasaulce(Allerton)11}, \cite{LeTreust(EmpiricalCoordination)14}, 
\cite{LarrousseLasaulceBloch(IT)14}. The objective is to characterize the set of target empirical distributions that are achievable by using a coding scheme. Optimal solutions have been characterized for lossless decoding \cite{LeTreust(CorrelationITW)14}, for state-dependent source and channel \cite{LeTreust(ISIT-TwoSided)15}, for channel feedback \cite{LeTreust(ISITfeedbacks)15}, for the two-agent case \cite{LarrousseLasaulceWigger(ITW)15}. Polar coding scheme for empirical coordination has been further  investigated in \cite{BlascoThobabenSkoglund12}, \cite{ChouBlochKliewer15}, \cite{ChouBlochKliewer16}, \cite{CerviaLuzziBlochLeTreust16}. In \cite{SchielerCuff(RateDistortion14)}, the authors measure the secrecy in  communication systems, using a rate-distortion approach that is closely related to empirical coordination.  In  \cite{LeTreustBloch(ISIT)16}, the authors investigate the connexion between the empirical coordination and the state-leakage induced by a coding scheme. Empirical coordination captures the knowledge of the transmitters,  regarding the random variables they don't observe.

The network topology of Fig. \ref{fig:StrategicEmpiricalCoordination} suggests that the decoder $P_2$ implements a strategy, knowing in advance the strategy of the encoder $P_1$. In contrast to the definition of the ``Nash Equilibrium'' \cite{Nash51},  this strategic interaction is not simultaneous and corresponds to the ``Stackelberg Equilibrium'' \cite{stackelberg-book-1934}. The transmission of strategic information has attracted a lot of attention in the literature of Game Theory \cite{CrawfordSobel1982StrategicInformation}, \cite{Forges94}. In \cite{KamenicaGentzkow11}, the authors investigate the problem of ``Bayesian Persuasion'' in which a sender wants to persuade a receiver to change her action. The state of the nature is a random variable observed by the sender $P_1$ but not by the receiver $P_2$. The sender applies a strategic quantification, designed in order to modify the posterior distributions of the receiver, regarding the state of the nature. The sender chooses an optimal signaling structure, knowing that the receiver implements a best-reply with respect to her posterior belief. 
This problem is called ``Information Design'' and relies on the `Splitting Lemma'' in the literature on Repeated Games with Incomplete Information \cite{AM95}, \cite{sorin-book-2002}, \cite{MertensSorinZamir15}. Dynamic versions of this problem were considered in \cite{Laraki01},  \cite{RenaultSolanVieille16}, \cite{ely2015beeps}, where the informed player announces her strategy in each stage. In \cite{JACKSONSONNENSCHEIN07}, the receiver fixes her strategy before the encoder, that react knowing in advance the strategy of the decoder. Strategic communication was considered more recently in the literature of Information Theory, for Gaussian source and channel with quadratic cost functions in \cite{AkyolLangbortBasar15}, \cite{AkyolLangbortBasar16} and \cite{SaritasYükselGezici16}.

In this paper, we characterize the encoding and decoding functions that form an equilibrium in the long-run game.  We introduce an auxiliary game in which both players choose the conditional distributions that maximize their expected utilities. The main result is stated in Sec. \ref{sec:StrategicEmpiricalCoordination}. In Sec. \ref{sec:InformationDesign}, we characterize the set of posterior distributions that are compatible with a rate-limited channel. In Sec. \ref{sec:ExMIMOFading}, we provide an example of non-aligned utility functions corresponding to parallel fading multiple access channels. Conclusion is stated in Sec. \ref{sec:conclusion} and the proofs of the main result are stated in App. \ref{sec:ProofThSEC2} and \ref{sec:ProofThSEC1}.


\section{Strategic Empirical Coordination}\label{sec:StrategicEmpiricalCoordination}

\subsection{Problem Statement}\label{sec:ProblemStatement}

We consider the problem of strategic empirical coordination depicted in Fig. \ref{fig:StrategicEmpiricalCoordination}. Notations $U^n$, $X^n$, $Y^n$,  $V^n$ stand for sequences of random variables of information source $u^n=(u_1,\ldots,u_n)\in\mc{U}^n$, inputs of the channel  $x^n\in\mc{X}^n$, outputs of the channel $y^n\in\mc{Y}^n$ and decoder's output $v^n\in\mc{V}^n$, respectively. The sets $\mc{U}$, $\mc{X}$, $\mc{Y}$, $\mc{V}$ have finite cardinality. The set of  probability distributions over $\mc{X}$ is denoted by $\Delta(\mc{X})$. The notation $||\QQ - \PP||_{1}= \sum_{x\in\mc{X}} |\QQ(x) - \PP(x)|$ stands for the $L_1$ distance between the probability distributions $\QQ$ and $\PP$. With a slight abuse of notation, we denote by $\QQ(x) \times \QQ(v|x) $, the product of distributions over $\Delta(\mc{X}\times \mc{V})$. Notation $Y  -\!\!\!\!\minuso\!\!\!\!-X    -\!\!\!\!\minuso\!\!\!\!-  U$ denotes the Markov chain property corresponding to $\PP(y|x,u) = \PP(y|x)$ for all $(u,x,y)$. Player $P_1$ observes a sequence of source symbols $u^n\in \mc{U}^n$ and chooses at random a sequence of channel inputs $x^n\in \mc{X}^n$. Player $P_2$ observes a sequence of channel outputs  $y^n\in \mc{Y}^n$ and chooses  at random  a sequence of actions $v^n\in \mc{V}^n$.

\begin{definition}[Strategies of both players]\label{def:Code}$\;$\\
$\bullet$ Player $P_1$ chooses a strategy $\sigma$ and player $P_2$ chooses a strategy $\tau$, defined as follows:
\begin{eqnarray}
&\sigma& : \mc{U}^{n} \longrightarrow \Delta(\mc{X}^n)  ,\label{eq:EncodingFunction}\\
&\tau& : \mc{Y}^n     \longrightarrow  \Delta( \mc{V}^n)  . \label{eq:DecodingFunction}
\end{eqnarray}
Both strategies $(\sigma,\tau)$ are stochastic.\\
$\bullet$ A pair of strategies $(\sigma,\tau)$ induces a joint probability distribution $\PP_{\sigma,\tau} \in\Delta(\mc{U}^{n} \times\mc{X}^{n}\times\mc{Y}^{n}  \times\mc{V}^{n} )$ over the $n$-sequences of symbols, defined by:
\begin{small}
\begin{eqnarray}
 \prod_{i=1}^n\PP\Big(U_i \Big) \times \PP_{\sigma} \Big(X^n\Big| U^n \Big)\times \prod_{i=1}^n \mc{T}\Big(Y_i \Big|X_i\Big) \times \PP_{\tau} \Big(V^n \Big| Y^n \Big).
\end{eqnarray}
\end{small}
\end{definition}

\begin{definition}[Expected $n$-stage utilities]\label{def:Utilities}$\;$\\
The utilities  of the $n$-stage game $\Phi_1^n$ and $\Phi_2^n$ are evaluated with respect to the marginal distribution $\PP_{\sigma,\tau}$ over the sequences $(U^n, V^n)$  and the utility functions $\phi_1(u,v)\in \R$, $\phi_2(u,v)\in \R$.
\begin{eqnarray}
\Phi_1^n(\sigma,\tau) &=& \E_{\sigma,\tau} \Bigg[ \frac{1}{n} \sum_{i=1}^n \phi_1(U_i,V_i) \Bigg] \nonumber\\
&=& \sum_{u^n,v^n}\PP_{\sigma,\tau}\Big(u^n,v^n \Big) \cdot  \Bigg[  \frac{1}{n} \sum_{i=1}^n \phi_1(u_i,v_i) \Bigg],\\
\Phi_2^n(\sigma,\tau) &=& \sum_{u^n,v^n}\PP_{\sigma,\tau}\Big(u^n,v^n \Big) \cdot  \Bigg[  \frac{1}{n} \sum_{i=1}^n \phi_2(u_i,v_i) \Bigg].
\end{eqnarray}
\end{definition}

\begin{definition}[Equilibrium utilities]\label{def:Equilibria}$\;$\\
We assume that player $P_2$ knows in advance the strategy $\sigma$ of player $P_1$ and chooses the mapping $\sigma\mapsto \tau(\sigma)$. $\big(\Phi_1^{\star}, \Phi_2^{\star}\big) \in \R^2$ is a pair of equilibrium utilities if there exists  strategies $\big(\sigma^{\star},\tau^{\star}(\sigma)\big)$ for both players $P_1$ and $P_2$ that satisfy:\\
1) $\Big(\Phi_1^n\big(\sigma^{\star},\tau^{\star}(\sigma)\big) , \Phi_2^n\big(\sigma^{\star},\tau^{\star}(\sigma)\big) \Big)$ converge to $\Big( \Phi_1^{\star} , \Phi_2^{\star} \Big) $, as $n \longrightarrow + \infty$,\\
2) for all $\varepsilon>0$, there exists a $\bar{n} \in \N$ such that for all $n\geq\bar{n}$, the two following equilibrium conditions are satisfied:
\begin{eqnarray}
\forall\sigma,\quad  \Phi_2^n\big(\sigma,\tau^{\star}(\sigma)\big) &\geq&\max_{\tilde{\tau}}\; \Phi_2^n\big(\sigma,\tilde{\tau}\big)  -  \varepsilon,\\
\Phi_1^n\big(\sigma^{\star},\tau^{\star}(\sigma^{\star})\big) &\geq&\max_{\tilde{\sigma} }\; \Phi_1^n\big(\tilde{\sigma},\tau^{\star}(\tilde{\sigma})\big)  -  \varepsilon.
\end{eqnarray}
\end{definition}

\begin{remark}
 In Definition \ref{def:Equilibria}, player $P_2$ has access to strategy $\sigma$ of player $P_1$, before choosing her strategy $\sigma\mapsto \tau(\sigma)$. This ``Stackelberg Equilibrium'' \cite{stackelberg-book-1934} hypothesis comes from the topology of the point-to-point network, Fig. \ref{fig:StrategicEmpiricalCoordination}.
\end{remark}



\subsection{Target Probability Distribution}\label{sec:TargetDistribution}

In  this section, we characterize the pair of equilibrium utilities $\big(\Phi_1^{\star}, \Phi_2^{\star}\big)$, by using a target probability distribution:
\begin{eqnarray}
\PP_{\sf{u}}(u) \times \QQ(v |u)  \in\Delta(\mc{U}   \times \mc{V}).\label{eq:CondDistribution}
\end{eqnarray}
Intuitively, the strategies $\sigma$ and $\tau$ of Definition \ref{def:Code} form a coding scheme. An auxiliary random variable $W$  is used to characterize the message $w^n \in \mc{W}^n$ sent by the encoder $P_1$ and the message $\hat{w}^n \in \mc{W}^n$ decoded by the decoder $P_2$. The decoding is correct if $w^n  = \hat{w}^n$. The objective is to control the empirical distribution $Q^n(u,w,v) \in\Delta(\mc{U} \times \mc{W}  \times \mc{V})$ of the sequences of actions and messages $(U^n,W^n,V^n)$, in order to achieve the following joint distribution:
\begin{eqnarray}
\PP_{\sf{u}}(u) \times \QQ(w | u)\times \QQ(v | w ) \in\Delta(\mc{U} \times \mc{W}  \times \mc{V}),\label{eq:JointDistribution}
\end{eqnarray}
with marginals on $(U,V)$ given by \eqref{eq:CondDistribution}. The auxiliary random variable $W$ captures the common information shared by both players $P_1$ and $P_2$. The distribution of \eqref{eq:JointDistribution} satisfies the Markov chain: $U  -\!\!\!\!\minuso\!\!\!\!- W  -\!\!\!\!\minuso\!\!\!\!- V$.  We introduce the sets of target joint probability distributions $\PP_{\sf{u}}(u) \times \QQ(w | u)  \times\QQ(v | w ) \in \Delta(\mc{U} \times \mc{W}\times \mc{V})$ that are achievable $\Q_0$ and that satisfy the best-reply condition $\Q_2$ for player $P_2$.

\begin{definition}[Achievable target distributions]\label{def:AchievableEmpiricalDistribution}$\;$\\
We define the set $\Q_0$ of joint probability distributions $ \PP_{\sf{u}}(u) \times \QQ(w | u) \times \QQ(v |w) $, that satisfy:
\begin{eqnarray}
\Q_0 &=& \bigg\{  \PP_{\sf{u}}(u) \times \QQ(w | u) \times \QQ(v |w),  \quad \text{s.t.}\nonumber \\
&& \quad \max_{\PP(x)} I( X; Y )  -   I( W ;U  ) \geq 0  \bigg\}.\label{eq:InfoConstraint0}
\end{eqnarray}
The set $\Q_0$  is convex since the mutual information $I( W ;U  )$ is convex in $\QQ(w | u) $, for fixed $\PP_{\sf{u}}(u)$.
\end{definition}

The information constraint  \eqref{eq:InfoConstraint0} of $\Q_0$ does not depend on the conditional distribution $\QQ(v |w)$, but only on the product of $\PP_{\sf{u}}(u) \times \QQ(w | u)$ and on the channel capacity $\max_{\PP(x)} I( X; Y )$. It ensures that the decoder can correctly recover the sequence $\hat{W}^n = W^n$, with high probability.

\begin{definition}[Strategic compatibility for player $P_2$]\label{def:Q2}$\;$\\
We define the set $\Q_2$  of joint  probability distributions $ \PP_{\sf{u}}(u) \times \QQ(w | u)\times   \QQ(v |w) $, that are strategically compatible for  $P_2$.
\begin{eqnarray}
\Q_2 &=& \bigg\{ \PP_{\sf{u}}(u) \times \QQ(w | u)\times   \QQ(v |w)    \quad \text{s.t.} , \nonumber \\
&&\qquad \E_{{\QQ}} \bigg[ \phi_2(U,V) \bigg] \geq  \E_{\widetilde{\QQ}} \bigg[ \phi_2(U,V) \bigg] , \nonumber \\
&&  \forall \;  \widetilde{\QQ}(u,w,{v}) =   \PP_{\sf{u}}(u) \times \QQ(w | u)  \times  \widetilde{\QQ}({v} | w)  \bigg\}.
\end{eqnarray}
We denote by $ \textsf{BR}_2\big(\QQ(w | u)\big)$, the set of distributions $\QQ(v |w)$ that are best-replies of player  $P_2$, for distribution $\QQ(w | u)$. The set $\Q_2$ is convex since the expectation is linear.\end{definition}

For all joint probability distributions $ \PP_{\sf{u}}(u) \times \QQ(w | u)$, the second player $P_2$ can generate a symbol $v \in \mc{V} $, by using another conditional probability distribution $  \widetilde{\QQ}({v} | w)$ than the prescribed one $\QQ(v |w) $. Definition \ref{def:Q2} ensures that  the target distribution is optimal for player $P_2$.

\begin{definition}[Set of target distributions]\label{def:Q}$\;$\\
We define the set $\Q$  of joint distributions $ \PP_{\sf{u}}(u) \times \QQ(v |u) $ that satisfy the following conditions:
\begin{eqnarray}
\Q &=& \bigg\{ \QQ(v |u)  \quad \text{ s.t. } \quad  \exists   W \text{ with } U  -\!\!\!\!\minuso\!\!\!\!- W  -\!\!\!\!\minuso\!\!\!\!- V, \nonumber \\
&& \sum_w \PP_{\sf{u}}(u) \times \QQ(w | u)\times   \QQ(v |w)   = \PP_{\sf{u}}(u) \times \QQ(v |u)\nonumber \\
&& \text{ and  }  \PP_{\sf{u}}(u) \times \QQ(w | u)\times   \QQ(v |w)  \in \Q_0 \cap \Q_2  \bigg\}.\end{eqnarray}
The set $\Q$ is convex since $\Q_0$ and $\Q_2$ are convex for any $W$. 
\end{definition}
The definition of $\Q$ involves an auxiliary random variable $W$ that satisfies the Markov chain $U  -\!\!\!\!\minuso\!\!\!\!- W  -\!\!\!\!\minuso\!\!\!\!- V$ and the marginal conditions $\PP_{\sf{u}}(u) \times \QQ(v |u)$. The random variable $W$ captures the information of $P_2$ regarding $U$ and the information of $P_1$ regarding $V$. The set $\Q$ characterizes the conditional distributions $\QQ(v|u)$ that are achievable  and that satisfy a best-reply condition  for player $P_2$. The conditional distributions $\QQ(v|u)$ outside $\Q$ cannot support equilibrium utilities.


\begin{theorem}\label{theo:StrategicEmpiricalCoordination}
The equilibrium utility $\Phi_1^{\star}$ of player $P_1$ is:
\begin{eqnarray}
\Phi_1^{\star}&=&  \max_{\QQ(v |u) \in \Q}\E\bigg[\Phi_1(U,V)\bigg]. \label{eq:hyp1}
\end{eqnarray}
The equilibrium utility $\Phi_2^{\star} =\E\big[\Phi_2(U,V)\big] $ is given by the expectation with respect to the conditional distribution $\QQ^{\star}(v |u) \in \Q$ that achieves the maximum in equation \eqref{eq:hyp1}.
\end{theorem}

The proof of Theorem \ref{theo:StrategicEmpiricalCoordination} is stated in App. \ref{sec:ProofThSEC2} and \ref{sec:ProofThSEC1}. The $n$-stage game of Definition \ref{def:Equilibria} is reformulated using a one-shot game in which $P_1$ chooses the optimal  achievable distribution $\QQ(w | u)$,  knowing that $P_2$ implements a best-reply $\QQ(v |w)\in \textsf{BR}_2\big(\QQ(w | u)\big)$.


\section{Information Design: Strategic Compression}\label{sec:InformationDesign}

\subsection{Control of the Posterior Distributions}\label{sec:ControlPosterior}

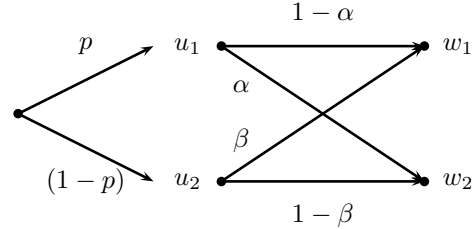
\begin{figure}[!ht]
\begin{center}
\psset{xunit=0.9cm,yunit=0.9cm}
\begin{pspicture}(-2.5,0)(3,3)
\rput(-1,0.5){$u_2$}
\rput(-1,2.5){$u_1$}
\rput(-2.5,0.5){$(1-p)$}
\rput(-2.5,2.5){$p$}
\psdots(-3.5,1.5)(-0.5, 0.5)(-0.5, 2.5)(2.5,0.5)(2.5,2.5)
\psline[linewidth=1pt]{->}(-3.5,1.5)(-1.5,2.5)
\psline[linewidth=1pt]{->}(-3.5,1.5)(-1.5,0.5)
\psline[linewidth=1pt]{->}(-0.5,0.5)(2.5,0.5)
\psline[linewidth=1pt]{->}(-0.5,2.5)(2.5,2.5)
\psline[linewidth=1pt]{->}(-0.5,0.5)(2.5,2.5)
\psline[linewidth=1pt]{->}(-0.5,2.5)(2.5,0.5)
\rput(3,0.5){$w_2$}
\rput(3,2.5){$w_1$}
\rput(1,3){$1 - \alpha$}
\rput(1,0){$1 - \beta$}
\rput(-0.2,1.9){$ \alpha$}
\rput(-0.2,1.1){$ \beta$}
\end{pspicture}
\caption{Signaling $\QQ(w|u)$ depending on $\alpha\in [0,1]$ and $\beta\in [0,1]$.  }
\label{fig:Signaling}
\end{center}
\end{figure}

We consider the binary information source $U \in \{u_1,u_2\}$ with parameter $p\in [0,1]$, $\PP_{\sf{u}}(u_1)=p$ and a binary auxiliary random variable $W \in \{w_1,w_2\}$.  The set of conditional distributions $\QQ(w|u)$ is represented by Fig. \ref{fig:Signaling} where $\QQ(w|u)$ involves two parameters $\alpha \in [0,1]$ and $\beta\in [0,1]$.
The posterior distributions of $u_1$ given $w_1$ or $w_2$ are denoted by $p_1$ and $p_2$ and write:
\begin{eqnarray}
\QQ(u_1 | w_1) &=& \frac{p \cdot (1-\alpha)}{p \cdot (1-\alpha) + (1-p) \cdot \beta } = p_1,\label{eq:posteriors1}\\
\QQ(u_1 | w_2) &=& \frac{p \cdot \alpha}{p \cdot \alpha + (1-p) \cdot (1-\beta) } = p_2.\label{eq:posteriors2}
\end{eqnarray}
By inverting the system of equations \eqref{eq:posteriors1} - \eqref{eq:posteriors2}, we obtain the parameters $(\alpha,\beta)$ corresponding to the pair of target posterior distributions $(p_1,p_2)$. 
\begin{eqnarray}
\alpha &=& \frac{p_2 \cdot (p_1 - p)}{p \cdot (p_1 - p_2)},\\
\beta &=&\frac{(1-p_1) \cdot (p - p_2)}{(1-p) \cdot (p_1 - p_2)}.
\end{eqnarray}

\begin{lemma}\label{lemma:AlphaBeta}
The parameters $(\alpha,\beta, p_1,p_2)$ correspond to Bernouilli distributions if and only if:\\
1) $p\notin \{0,1\}$ and\\
2) $p_1< p< p_2$ or $p_2< p< p_1$.
\end{lemma}


For each pair of posterior distributions $(p_1,p_2) \in[0,p[\times ]p,1]$, there exists a pair of $(\alpha,\beta)$ such that the prior distribution $p$ can be splitted according to  $p_1 \in [0,p[$ and $p_2 \in ]p,1]$.  

\subsection{One-Shot Transmission with a Noisy Channel}\label{sec:OneShotTransmission}

\begin{figure}[!ht]
\begin{center}
\psset{xunit=0.7cm,yunit=0.7cm}
\begin{pspicture}(-2,-4)(7,3)
\rput(-1,0.5){$u_2$}
\rput(-1,2.5){$u_1$}
\rput(-2.5,0.5){$(1-p)$}
\rput(-2.5,2.5){$p$}
\psdots(-3.5,1.5)(-0.5, 0.5)(-0.5, 2.5)(2.5,0.5)(2.5,2.5)
\psline[linewidth=1pt]{->}(-3.5,1.5)(-1.5,2.5)
\psline[linewidth=1pt]{->}(-3.5,1.5)(-1.5,0.5)
\psline[linewidth=1pt]{->}(-0.5,0.5)(2.5,0.5)
\psline[linewidth=1pt]{->}(-0.5,2.5)(2.5,2.5)
\psline[linewidth=1pt]{->}(-0.5,0.5)(2.5,2.5)
\psline[linewidth=1pt]{->}(-0.5,2.5)(2.5,0.5)
\rput(3,0.5){$w_2$}
\rput(3,2.5){$w_1$}
\rput(4,0.5){$x_2$}
\rput(4,2.5){$x_1$}
\psdots(4.5, 0.5)(4.5, 2.5)(7.5,0.5)(7.5,2.5)
\rput(8,0.5){$y_2$}
\rput(8,2.5){$y_1$}
\psline[linewidth=1pt]{->}(4.5,0.5)(7.5,0.5)
\psline[linewidth=1pt]{->}(4.5,2.5)(7.5,2.5)
\psline[linewidth=1pt]{->}(4.5,0.5)(7.5,2.5)
\psline[linewidth=1pt]{->}(4.5,2.5)(7.5,0.5)
\rput(6,3){$1 - \varepsilon$}
\rput(6,0){$1 - \varepsilon$}
\rput(4.8,1.9){$ \varepsilon$}
\rput(4.8,1.1){$ \varepsilon$}
\rput(1,3){$1 - \alpha$}
\rput(1,0){$1 - \beta$}
\rput(-0.2,1.9){$ \alpha$}
\rput(-0.2,1.1){$ \beta$}
\rput(-1,-3.5){$u_2$}
\rput(-1,-1.5){$u_1$}
\rput(-2.5,-3.5){$(1-p)$}
\rput(-2.5,-1.5){$p$}
\psdots(-3.5,-2.5)(-0.5, -3.5)(-0.5, -1.5)(7.5,-3.5)(7.5,-1.5)
\psline[linewidth=1pt]{->}(-3.5,-2.5)(-1.5,-1.5)
\psline[linewidth=1pt]{->}(-3.5,-2.5)(-1.5,-3.5)
\psline[linewidth=1pt]{->}(-0.5,-3.5)(7.5,-3.5)
\psline[linewidth=1pt]{->}(-0.5,-1.5)(7.5,-1.5)
\psline[linewidth=1pt]{->}(-0.5,-3.5)(7.5,-1.5)
\psline[linewidth=1pt]{->}(-0.5,-1.5)(7.5,-3.5)
\rput(8,-3.5){$y_2$}
\rput(8,-1.5){$y_1$}
\rput(3.5,-1){$(1 - \alpha) \star \varepsilon$}
\rput(3.5,-4){$(1 - \beta) \star \varepsilon$}
\rput(0.2,-2.1){$ \alpha \star \varepsilon$}
\rput(0.2,-2.9){$ \beta \star \varepsilon$}
\end{pspicture}
\caption{The concatenation of the conditional distributions $\QQ(w|u)$ and $\mc{T}(y|x)$ can be expressed using a binary symmetric channel $\PP(y|u)$, with noise parameters $ \alpha \star \varepsilon\in [0,1]$ and $ \beta \star \varepsilon\in [0,1]$.}
\label{fig:NoisyChannel}
\end{center}
\end{figure}
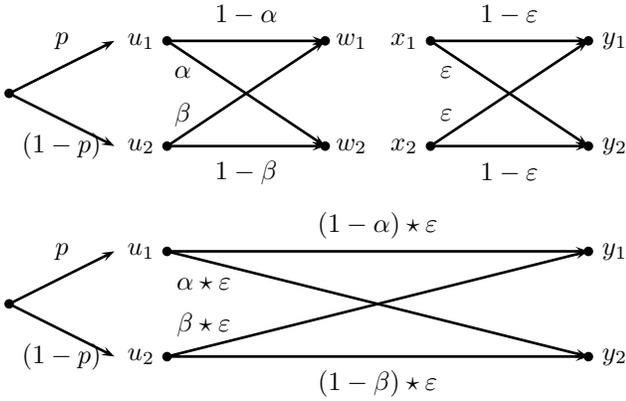

We consider a binary symmetric channel $\mc{T}(y|x)$ with noise parameter $\varepsilon \in [0,0.5]$, as represented by Fig. \ref{fig:NoisyChannel}. 
The concatenation of the signaling structure $\QQ(w|u)$ with the noisy channel $\mc{T}(y|x)$ can be directly expressed with parameters $ \alpha \star \varepsilon \in [0,1]$ and $ \beta \star \varepsilon\in [0,1] $, instead of $ \alpha$, $ \beta$ and $ \varepsilon$.
\begin{eqnarray}
\alpha \star \varepsilon &=&  (1 - \alpha) \cdot  \varepsilon +  \alpha \cdot (1-  \varepsilon). \label{eq:convolution}\end{eqnarray}

\begin{proposition}\label{prop:Alpha}
A pair of posterior distributions $(p_1,p_2)$ is achievable with the noisy channel $\mc{T}(y|x)$ if and only if there exists $(\alpha,\beta)$ such that: 
\begin{eqnarray}
\alpha \star \varepsilon &=& \frac{p_2 \cdot (p_1 - p)}{p \cdot (p_1 - p_2)},\label{eq:CoroAplha1}\\
\beta \star \varepsilon &=&\frac{(1-p_1) \cdot (p - p_2)}{(1-p) \cdot (p_1 - p_2)}.\label{eq:CoroAplha2}
\end{eqnarray}
\end{proposition}

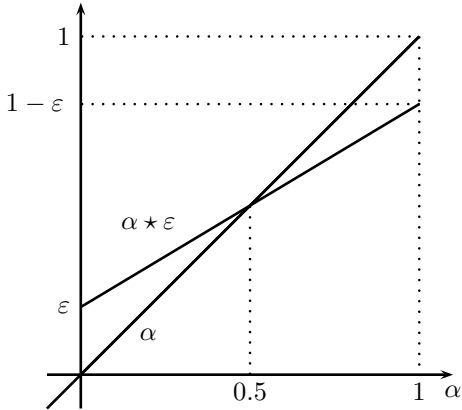
\begin{figure}[!ht]
\begin{center}
\psset{xunit=4.5cm,yunit=4.5cm}
\begin{pspicture}(-0.1,-0.1)(1,1.1)
\psline[linewidth=1pt]{->}(-0.1,0)(1.1,0)
\psline[linewidth=1pt]{->}(0,-0.1)(0,1.1)
\psline[linewidth=1pt](-0.1,-0.1)(1,1)
\psline[linewidth=1pt](0,0)(1,1)
\psline[linewidth=1pt](0,0.2)(1,0.8)
\psline[linewidth=1pt,linestyle=dotted](0.5,0.5)(0.5,0)
\psline[linewidth=1pt,linestyle=dotted](0,1)(1,1)(1,0)
\psline[linewidth=1pt,linestyle=dotted](0,0.8)(1,0.8)
\rput(-0.05,0.2){$ \varepsilon$}
\rput[r](-0.05,0.8){$1-  \varepsilon$}
\rput(0.2,0.45){$\alpha \star  \varepsilon$}
\rput(0.2,0.12){$\alpha $}
\rput(-0.05,1){$1$}
\rput(1.1,-0.05){$\alpha $}
\rput(1,-0.05){$1$}
\rput(0.5,-0.05){$0.5$}
\end{pspicture}
\caption{Function $\alpha \star \varepsilon$ depending on $\alpha \in[0,1]$. }
\label{fig:AlphaStarEpsilon}
\end{center}
\end{figure}

We can see on Fig. \ref{fig:AlphaStarEpsilon} that no $\alpha$ exists such that $\alpha \star \varepsilon  =  \alpha \cdot (1-  2\varepsilon) + \varepsilon > 1- \varepsilon$ or  $\alpha \star \varepsilon < \varepsilon$. This imposes a restriction over the set of achievable posteriors $(p_1,p_2)$, that is represented by the  ``region of the circle'', in Fig. \ref{fig:ShannonCaseB_2016_06_10}.

\vspace{-0.3cm}

\subsection{Block Transmission with a Noisy Channel}\label{sec:BlockTransmission}
\begin{figure}[ht!]
\includegraphics[width=0.53\textwidth]{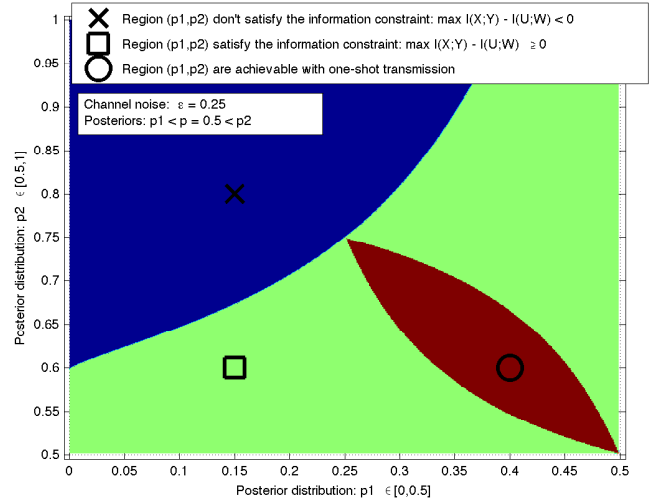}
\caption{Regions of achievable posteriors $(p_1,p_2)$, for channel noise $\varepsilon = 0.25$.}\label{fig:ShannonCaseB_2016_06_10}
\end{figure}

We consider the scenario represented by Fig. \ref{fig:StrategicEmpiricalCoordination} where the symbols are encoded by blocks. Theorem \ref{theo:StrategicEmpiricalCoordination} states that the conditional distribution $\QQ(w|u)$ defined with $(\alpha,\beta)$ in Fig. \ref{fig:Signaling}, is achievable if and only if:
\begin{eqnarray}
&&\max_{\PP(x)} I(X;Y) - I(U;W) \geq  0,\\
&\Longleftrightarrow &1 - H(\varepsilon) - H\Big(\PP(W_1)\Big) \nonumber \\
&&+ \;\; p \cdot H(\alpha ) + (1-p) \cdot H(\beta) \geq  0,\\
&&\text{ with } \PP(W_1) = p \cdot (1-\alpha) + (1-p ) \cdot \beta. \nonumber
\end{eqnarray}
Fig. \ref{fig:ShannonCaseB_2016_06_10} represents three regions of posterior distributions $(p_1,p_2)$.  The ``region of the square'' corresponds to posteriors $(p_1,p_2)$ that satisfy the information constraint of the set $\Q_0$. It includes the ``region of the circle'', in which the posteriors $(p_1,p_2)$  are achievable in one-shot and satisfy equations \eqref{eq:CoroAplha1} - \eqref{eq:CoroAplha2} of Proposition \ref{prop:Alpha}. The posteriors $(p_1,p_2)$ that belong to the ``region of the cross'', are not achievable.

\section{Non-Aligned Utilities: Parallel Fading Multiple Access Channels}\label{sec:ExMIMOFading}

\subsection{Power Allocation Game}\label{sec:PowerAllocationGame}


In this section, we consider an example involving two transmitters that communicate with two base stations over parallel fading multiple access channels \cite{Belmega-TWC-2009}. The transmissions are simultaneous and cause mutual interferences. Both transmitters have maximal power equal to 1 and the noise variance is $\sigma^2 =1$. We consider two possible configurations for the random channel gains $G \in \{g_A, g_B\}$, described by the following table and chosen with probability $(p,1-p)$ for $p=0.5$. In this section, the channel gains are considered as an information source $U=G$. 
\begin{center}
\begin{tabular}{|c|c|c|}
\hline
& $g_A$  & $g_B$ \\    
        \hline
  $g_{11}$ & 1.1878 &  0.1811\\
      \hline
  $g_{12}$ & 1.1566&  1.4475\\
    \hline
  $g_{21}$ &   0.8407  &  0.0717\\
      \hline
  $g_{22}$ & 0.6293& 0.6858\\
    \hline
 \end{tabular}
\end{center}
The notation $g_{12}$ corresponds to the channel gain between the first transmitter and the second base station. We assume that the power allocation of the first transmitter is fixed $(a_1, 1-a_1) = (0.16,0.84)$. The second transmitter chooses a power allocation $(v,1-v)$ from the discrete set $v \in \mc{V} = \{0,0.25,0.5,0.75,1\}$, in order to maximize her expected utility $\E\big[\phi_2(G,v)\big]$. 
\begin{eqnarray}
\E\big[\phi_2(G,v)\big]   &=& p \cdot  \phi_2(g_A,v) +  (1-p) \cdot  \phi_2(g_B,v),\\
\phi_2(g,v) &=&  \log_2\bigg( 1 + \frac{v \cdot g_{21}}{\sigma^2 + a_1\cdot g_{11}}\bigg) \nonumber \\
&+& \log_2\bigg( 1 + \frac{(1-v)\cdot g_{22}}{\sigma^2 + (1-a_1)\cdot g_{12}}\bigg).
\end{eqnarray}
\begin{figure}[ht!]
\includegraphics[width=0.5\textwidth]{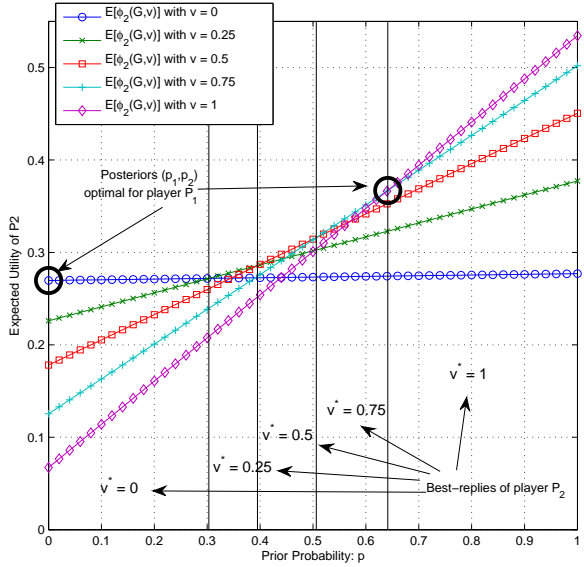}
\caption{Expected utility function $\E\big[\phi_2(G,v)\big]$ of the second transmitter $P_2$, depending on her power allocation $v \in  \{0,0.25,0.5,0.75,1\}$. The best-reply of $P_2$ is denoted by $v^{\star}$ and  depends on her prior probability $p\in[0,1]$. }\label{UtilityBSandT2_2016_06_17}
\end{figure}
We consider the game between the base station $P_1$ and the second transmitter $P_2$. The base station is informed of the realization of the channel gains $G\in\{g_A,g_B\}$ and wishes to persuade the second transmitter $P_2$ to choose a favorable power allocation $v\in \mc{V}$. The utility of the base station $\phi_{1}(G,v)$ is equal  to the utility of the first transmitter:
\begin{eqnarray}
\phi_{1}(g,v) &=&  \log_2\bigg( 1 + \frac{a_1 \cdot g_{11}}{\sigma^2 + v \cdot g_{12}}\bigg) \nonumber \\
&+& \log_2\bigg( 1 + \frac{(1-a_1) \cdot g_{12}}{\sigma^2 + (1-v)\cdot g_{22}}\bigg).
\end{eqnarray}
 \begin{figure}[h!]
\includegraphics[width=0.5\textwidth]{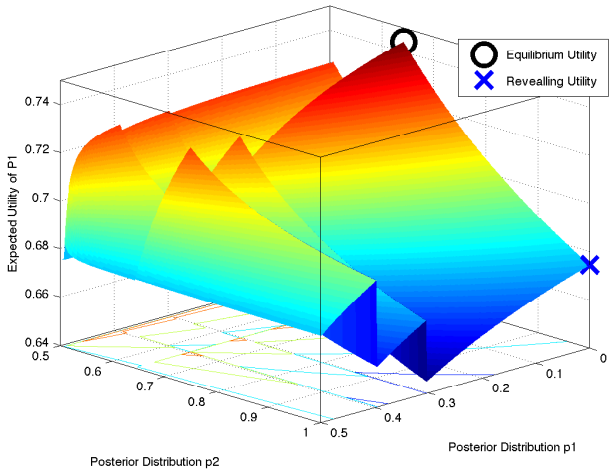}
\caption{Expected utility $\E\big[\phi_1(G,v)\big] $ depending on the pair of posterior distributions $(p_1,p_2)$, induced by the signaling $\QQ(w|u)$ of Fig. \ref{fig:Signaling}. The discontinuities are due to the changes of best-reply $v^{\star}$ of $P_2$, see Fig. \ref{UtilityBSandT2_2016_06_17}. }\label{MIMOStrategicTA1_2016_06_16}
\end{figure}
Depending on the realization of the channel gains $g_A$ or $g_B$, the base station $P_1$ sends a strategic signal $W \in \{w_1,w_2\}$ to the second transmitter $P_2$ using the signaling structure $\QQ(w|u)$, depicted in Fig. \ref{fig:Signaling}. Given the strategic signal $W$, player $P_2$ chooses the power allocation $v \in  \{0,0.25,0.5,0.75,1\}$  that maximizes her own expected utility $\E\big[\phi_{2}(G,v)\big]$. 

Fig. \ref{UtilityBSandT2_2016_06_17} represents the expected utility function $\E\big[\phi_2(G,v)\big]$ depending on the prior probability $p\in [0,1]$, for each power allocation $v \in  \{0,0.25,0.5,0.75,1\}$. The best-reply allocation of $P_2$ is denoted by $v^{\star}$ and depends on the interval to which belongs the prior probability $p\in [0,1]$.
Upon receiving symbol $w\in\mc{W}$ and knowing the joint distribution $\PP_{\sf{u}}(u) \times \QQ(w | u)$ of Fig. \ref{fig:Signaling}, player $P_2$ implements a best-reply  $\QQ(v |w)\in \textsf{BR}_2\big(\QQ(w | u)\big)$. Theorem \ref{theo:StrategicEmpiricalCoordination} guarantees that the i.i.d. distribution $\QQ(v |w)$ induces an $\varepsilon$-best-reply in the long-run game.

\begin{figure}[ht!]
\includegraphics[width=0.45\textwidth]{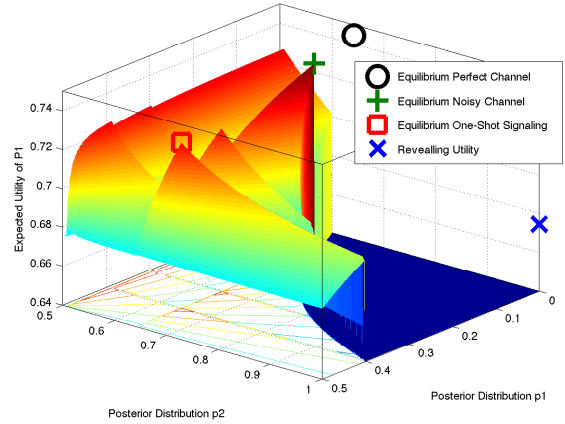}
\caption{The  ``green plus'' represents the equilibrium utility $\Phi_1^{\star}=  \max_{\QQ(v |u)\in \Q} \E\big[\phi_1(G,v)\big] $ characterized by Theorem \ref{theo:StrategicEmpiricalCoordination}, for channel noise  $\varepsilon = 0.25$.}\label{UtilityP1C_2016_07_06}
\end{figure} 

The base station $P_1$ already knows that $P_2$ implements a best-reply. It chooses accordingly the signaling structure $\QQ(w|u)$ that induces the more favorable response of $P_2$. 
Fig \ref{MIMOStrategicTA1_2016_06_16} shows the expected utility $\E\big[\phi_{1}(G,v)\big]$ of $P_1$, depending on the pair of posterior distributions $(p_1,p_2)$, induced by the signaling structure $\QQ(w|u)$. The discontinuities are due to the changes of best-reply $v^{\star}$ of $P_2$, according to the posterior distributions $(p_1,p_2)$, see Fig. \ref{UtilityBSandT2_2016_06_17}. The equilibrium utility $\E[ \phi_{1}] \simeq 0.74$ of $P_1$ is represented by the ``black circle'' and corresponds to the posterior distributions $(p_1,p_2) =(0,0.6415)$, with $(\alpha,\beta) = (1, 0.4424)$. This pair of optimal posteriors for $P_1$ is also represented by the ``black circles'', on Fig. \ref{UtilityBSandT2_2016_06_17}. This equilibrium utility provides $9.1\%$ of improvement compared to the revealing strategy (``blue cross'' $\E[ \phi_{1}] \simeq 0.67$), \textit{i.e.} when the channel gains are revealed to the second transmitter $P_2$, with $\alpha=\beta =0$. The authors would like to thank Claudio Weidmann for fruitful discussions regarding this section.


\subsection{Rate-Limited Channel between Players $P_1$ and $P_2$}\label{sec:NoisyChannel}

We assume that the channel between the base station $P_1$ and the second transmitter $P_2$ is rate-limited, \textit{i.e.} there is a noisy channel $\mc{T}(y|x)$ between $P_1$ and $P_2$, as depicted in Fig. \ref{fig:StrategicEmpiricalCoordination}. 
 \begin{figure}[ht!]
\includegraphics[width=0.45\textwidth]{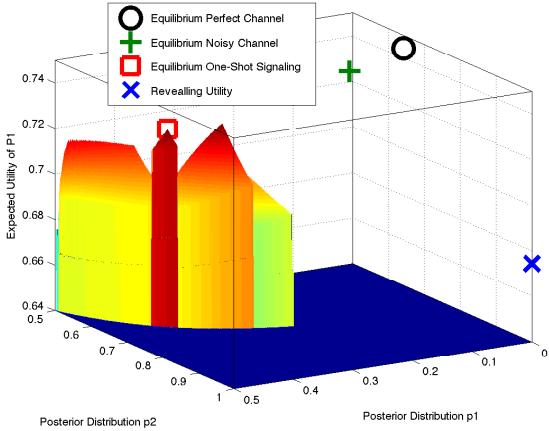}
\caption{The ``red square'' represents the equilibrium utility for one-shot transmission over a noisy channel that satisfies the conditions of Prop. \ref{prop:Alpha}.}\label{UtilityP1B_2016_07_06}
\end{figure}
The rate-limited constraint reduces the set of posterior distributions $(p_1,p_2)$, see Fig. \ref{fig:ShannonCaseB_2016_06_10}. We determine the equilibrium solutions for a binary symmetric channel with parameter $\varepsilon=0.25$, for one-shot and block transmission, as depicted in Fig. \ref{UtilityP1B_2016_07_06} and Fig. \ref{UtilityP1C_2016_07_06}. These figures correspond to the utility $\E\big[\phi_1(G,v)\big]$ of Fig. \ref{MIMOStrategicTA1_2016_06_16}, restricted to the ``region of the circle'' and ``region of the square'', in Fig. \ref{fig:ShannonCaseB_2016_06_10}. The equilibrium utilities correspond respectively to $\E\big[\phi_1(G,v)\big] \simeq 0.72$ and $\E\big[\phi_1(G,v)\big] \simeq 0.73$. In both cases, the revealing strategy is not achievable since it is not compatible with the rate-limitation imposed by the noisy channel $\mc{T}(y|x)$.



\section{Conclusion}\label{sec:conclusion}

We investigate the strategic coordination of an encoder and a decoder, endowed with non-aligned utility functions. We characterize the encoding and decoding functions that form an equilibrium, by using empirical coordination. The equilibrium solution is related to an auxiliary game in which both players choose the conditional distributions in order to maximize their expected utilities. We characterize the set of posterior distributions that are compatible with a rate-limited channel between the encoder and the decoder and we provide an example of non-aligned utility functions corresponding to parallel fading multiple access channels.



\appendices

\section{Proof of the Upper Bound for Theorem \ref{theo:StrategicEmpiricalCoordination}}\label{sec:ProofThSEC2}

In this section, we consider that the strategies $(\sigma^{\star},\tau^{\star})$ form an equilibrium. $Q^n(u,v)$ denotes the empirical distribution of the sequences $(u^n,v^n)$.
\begin{eqnarray}
&&\Phi_1^n(\sigma^{\star},\tau^{\star}) \nonumber \\
&=& \sum_{u^n,x^n,\atop y^n,v^n} \PP_{\sigma^{\star}\tau^{\star} } (u^n,x^n,y^n,v^n)  \cdot \frac{1}{n} \sum_{i=1}^n \phi_1(u_i,v_i) \nonumber\\
&=& \sum_{u^n,x^n,\atop y^n,v^n} \PP_{\sigma^{\star}\tau^{\star} } (u^n,x^n,y^n,v^n)  \cdot \sum_{u,v} Q^n(u,v) \cdot \phi_1(u,v)\nonumber\\
&=& \sum_{u,v}  \E_{\sigma^{\star},\tau^{\star} } \bigg[Q^n(u,v) \bigg]\cdot  \phi_1(u,v)    \label{eq:MaxUtility1}
\end{eqnarray}
We denote by $\QQ^{\star}(u,v)$, the expected empirical distribution $ \E_{\sigma^{\star},\tau^{\star} } \big[Q^n(u,v) \big] =\QQ^{\star}(u,v)$ corresponding to the strategies $(\sigma^{\star},\tau^{\star})$. The marginal distribution over $U$ satisfies $\sum_{v\in \mc{V}} \E_{\sigma^{\star},\tau^{\star} } \big[Q^n(u,v) \big]  = \PP_{\sf{u}}(u)$. We denote by $\Q^c$ the complementary of the set $\Q$. We show that the conditional distribution $\QQ^{\star}(v|u)  \in \Q$ should belong to the set $\Q$ of Definition \ref{def:Q}. Player $P_1$ cannot choose a distribution $\QQ(v|u) \in \Q^c$ that lie outside $\Q$, in order to maximize her long-run utility: 
\begin{eqnarray}
\Phi_1^n(\sigma^{\star},\tau^{\star})  &\leq &  \max_{\QQ(v |u) \in \Q}\E\bigg[\Phi_1(U,V)\bigg].
\end{eqnarray}


\subsection{Information Constraint}\label{sec:ProofInfoConstraint}
We show that there exists an auxiliary random variable $W$ that satisfies the Markov chain $U  -\!\!\!\!\minuso\!\!\!\!- W  -\!\!\!\!\minuso\!\!\!\!- V$ and the information constraint $\max_{\PP(x)}I(X;Y) - I(U;W)\geq0$ of Definition \ref{def:AchievableEmpiricalDistribution}. 
\begin{eqnarray}
0 &\leq& I(X^n ; Y^n) - I(U^n;Y^n) \label{eq:Converse1} \\
&\leq& \sum_{i=1}^n H( Y_i)  -  \sum_{i=1}^n H(Y_i | X_i) \nonumber \\
&-& \sum_{i=1}^n  H(U_i) + \sum_{i=1}^nH(U_i|Y^{n},U^{i-1})  \label{eq:Converse4} \\
&=& \sum_{i=1}^n I(X_i ; Y_i)  -\sum_{i=1}^n I(U_i; W_i)  \label{eq:Converse6} \\
&\leq& n \cdot \max_{\PP(x)} I(X ; Y)  -\sum_{i=1}^n I(U_i; W_i)  \label{eq:Converse7} \\
&=& n \cdot  \bigg(\max_{\PP(x)} I(X ; Y)  -  I(U ; W_T , T) \bigg) \label{eq:Converse9} \\
&=& n \cdot  \bigg(\max_{\PP(x)} I(X ; Y)  -  I(U ; W) \bigg) \label{eq:Converse10} .
\end{eqnarray}
Equation \eqref{eq:Converse1} comes from the Markov chain $Y^n  -\!\!\!\!\minuso\!\!\!\!- X^n   -\!\!\!\!\minuso\!\!\!\!- U^n$.\\
Equation \eqref{eq:Converse4} comes from the memoryless property of the channel and the i.i.d. property of the source.\\
Equation \eqref{eq:Converse6} comes from the identification of the auxiliary random variable $W_i = (Y^{n},U^{i-1})$ that satisfies the Markov chain of the set $\Q$ for all $i\in \{1,\ldots,n\}$:
\begin{eqnarray}
&&U_i -\!\!\!\!\minuso\!\!\!\!- W_i  -\!\!\!\!\minuso\!\!\!\!- V_i. \label{eq:MarkovConverse1}
\end{eqnarray}
Equation \eqref{eq:Converse7} comes from taking the maximum over $\PP(x)$.\\
Equation \eqref{eq:Converse9} comes from the introduction of the uniform random variable $T$ over the indices $\{1,\ldots,n\}$,  the independence between $T$ and $U_T$, that implies $I(T;U_T)=0$ and the i.i.d. property of the source $U_T =U$.\\
Equation \eqref{eq:Converse10} comes from the identification of the auxiliary random variable $ W=(W_T,T) = (Y^{n},U^{T-1}, T)$ and the Markov chain property:
\begin{eqnarray}
&&U -\!\!\!\!\minuso\!\!\!\!- (W_T,T)  -\!\!\!\!\minuso\!\!\!\!- V_T.\label{eq:MarkovConverse2}
\end{eqnarray}

Hence, the joint distribution $\PP_{\sf{u}}(u) \times \QQ^{\star}(w | u) \times \QQ^{\star}(v|w)$ induced by the auxiliary random variable $ W=(W_T,T) = (Y^{n},U^{T-1}, T)$ satisfies:\\
1) the Markov chain condition $U -\!\!\!\!\minuso\!\!\!\!- W  -\!\!\!\!\minuso\!\!\!\!- V$,\\
2) the marginal condition $\sum_{w} \PP_{\sf{u}}(u) \times \QQ^{\star}(w | u) \times \QQ^{\star}(v|w) = \PP_{\sf{u}}(u) \times \QQ^{\star}(v | u)$ given by equation \eqref{eq:MaxUtility1},\\
3) the positive information constraint  \eqref{eq:Converse10}. \\
In App. \ref{sec:ProofprofitableDeviationQ2}, we prove that the distribution $\PP_{\sf{u}}(u) \times \QQ^{\star}(w | u) \times \QQ^{\star}(v|w)$ belongs to the set $\Q_2$, \textit{i.e.} it satisfies the best-reply condition for $P_2$.



\subsection{Profitable Deviation of the Second Player}\label{sec:ProofprofitableDeviationQ2}

In this section, we consider a target distribution $\PP_{\sf{u}}(u) \times {\QQ}(w | u)  \times \QQ(v|w) \in \Q_0 \cap \Q_2^c$, that lies outside $\Q_2$ and that is achievable, \textit{i.e.} $\max_{\PP(x)} I(X ; Y) - I(W;U)\geq0$. Hence, there exists a distribution $ \widetilde{\QQ}(v|w) \neq {\QQ}(v|w)$ that increases the utility of player $P_2$:
\begin{eqnarray}
&& \E_{\widetilde{\QQ}}  \bigg[ \phi_2(U,V) \bigg] >  \E_{{\QQ}}
\bigg[ \phi_2(U,V) \bigg] , \nonumber \\
&\Longleftrightarrow&\sum_{u,w,v}\PP_{\sf{u}}(u) \times {\QQ}(w | u)  \times \widetilde{\QQ}(v|w) \times\phi_2(u,v) \nonumber \\
&>& \sum_{u,w,v}\PP_{\sf{u}}(u) \times {\QQ}(w | u)  \times \QQ(v|w) \times\phi_2(u,v).
\end{eqnarray}
Since the  target distribution $\PP_{\sf{u}}(u) \times {\QQ}(w | u)  \times \QQ(v|w) \in \Q_0$ is achievable, the second player  $P_2$ correctly decodes the sequence $W^n$, with high probability. There exists a deviating strategy $\tau\neq\tau^{\star}$ based on the i.i.d. distribution $ \widetilde{\QQ}(v|w)$ that is profitable for $P_2$. Hence there is a $\delta>0$ such that the equilibrium condition is not satisfied:
\begin{eqnarray}
\Phi_2^n(\sigma^{\star},\tau) &>& \Phi_2^{\star}  + \delta.
\end{eqnarray}


\section{Proof of the Lower Bound for Theorem \ref{theo:StrategicEmpiricalCoordination}}\label{sec:ProofThSEC1}

In this section, we provide a coding scheme $(\sigma^{\star},\tau^{\star})$ that satisfies both equilibrium conditions of Definition \ref{def:Equilibria}:
\begin{eqnarray}
\Phi_2^n\big(\sigma^{\star},\tau^{\star}(\sigma^{\star})\big) &\geq&\max_{\tilde{\tau}}\; \Phi_2^n\big(\sigma^{\star},\tilde{\tau}\big)  -  \varepsilon,\\
\Phi_1^n\big(\sigma^{\star},\tau^{\star}(\sigma^{\star})\big) &\geq&\max_{\tilde{\sigma} }\; \Phi_1^n\big(\tilde{\sigma},\tau^{\star}(\tilde{\sigma})\big)  -  \varepsilon.
\end{eqnarray}


\subsection{Separated Source-Channel Coding}\label{sec:ExistenceOptimalStrategy}

We consider the target joint probability distribution $\QQ(u,w,v) = \PP_{\sf{u}}(u) \times \QQ(w | u)\times  \QQ(v |w) \in   \Q_0 \cap \Q_2$ that corresponds to the optimal utility $\Phi_1^{\star}$ of equation \eqref{eq:hyp1}. By definition of $\Q_0$, the target distribution $\QQ(u,w,v)$ satisfies the information constraint:
\begin{eqnarray}
\max_{\PP(x)} I( X; Y )  -   I( W ;U  ) \geq 0. \label{eq:InfoConstraint}
\end{eqnarray}
In this section, we assume that equation \eqref{eq:InfoConstraint} is satisfied with strict inequality \eqref{eq:InfoConstraintS}. The case of equality in the information constraint will be treated in App. \ref{sec:EqualityIC}. 
\begin{eqnarray}
\max_{\PP(x)} I( X; Y )  -   I( W ;U  ) > 0. \label{eq:InfoConstraintS}
\end{eqnarray}
Inequality \eqref{eq:InfoConstraintS} implies that there exists a small parameter $\delta>0$ and a rate  $\textsf{R}\geq 0 $, such that:
\begin{eqnarray}
\textsf{R}  &\geq&       I( W ; U )  + \delta  \label{eq:AchievabilitySEC1} , \\
\textsf{R}  &=&   \max_{\PP(x)} I( X; Y )  -  \delta  \label{eq:AchievabilitySEC2}  .
\end{eqnarray}
We define a code $c = (f,g)\in \mc{C}(n)$ using the following encoding and decoding functions, that involve  sequence $W^n$:
 \begin{eqnarray}
&f& : \mc{U}^{n} \longrightarrow  \mc{X}^n \times \mc{W}^n  ,\label{eq:EncodingFunctionC}\\
&g& : \mc{Y}^n     \longrightarrow   \mc{W}^n  \times \mc{V}^n. \label{eq:DecodingFunctionC}
\end{eqnarray}
We show the existence of an optimal code $c^{\star} \in\mc{C}(n)$ such that the empirical distribution  $Q^{n }(u,w,v)$ of symbols $(U^{n},W^{n },V^{n })$, is close to the target distribution $\QQ(u,w,v)  =  \PP_{\sf{u}}(u) \times \QQ(w | u) \times \QQ(v |w)$, with large probability. More precisely, we prove that for all $\varepsilon>0$, there exists an $\bar{n}\in \N$, such that for all $n\geq \bar{n}$, there exists a code $c^{\star} \in \mc{C}(n)$ that satisfies:
\begin{eqnarray}
\PP_{c^{\star} } \bigg(\Big|\Big|Q^{n }(u,w,v) - \QQ(u,w,v) \Big|\Big|_{1} \geq \varepsilon\bigg)\leq \varepsilon.
\end{eqnarray}
The parameter $\varepsilon>0$ is involved in both the definition of the typical sequences and the upper bound of the error probability.

\begin{itemize}
\item[$\bullet$] \textit{Random codebook.} We generate $| \mc{M}   |= 2^{n   \sf{R}   } $ sequences $W^n(m)$ and $X^n(m)$,  drawn from the marginal i.i.d. probability distributions $\QQ_{\sf{w}}^{\times n} $ and $\PP_{\sf{x}}^{\times n} $ with index  $m\in \mc{M} $. 
\item[$\bullet$] \textit{Encoding function.} The encoder observes the sequence of symbols of source $U^n \in  \mc{U}^n$. It finds an index $m\in \mc{M}$ such that the sequences  $\big(U^n,W^n(m)\big) \in A_{\varepsilon}^{{\star}{n}} (\QQ)$ are jointly typical. The encoder sends the sequence $X^n(m)$ corresponding to the index $m\in \mc{M}$.
\item[$\bullet$] \textit{Decoding function.} The decoder observes the sequence of channel output $Y^n\in\mc{Y}^n$. It returns an index $\hat{m}\in \mc{M}$ such that the sequences  $\big(Y^n,X^n(\hat{m})\big) \in A_{\varepsilon}^{{\star}{n}} (\QQ)$ are jointly typical. It deduces the sequence $W^n(\hat{m})$ and returns $V^n $, drawn from the conditional probability distribution $\QQ_{\sf{v|w}}^{\times n}$  depending on $W^n(\hat{m})$.  
\item[$\bullet$] \textit{Error Event.} An error occurs in the coding process if: 1) the indexes $m\in \mc{M}$ and $\hat{m}\in \mc{M}$ are not equal, not unique or does not exists, 2) the sequences of symbols $\big( U^n ,W^n,V^n \big) \notin A_{\varepsilon}^{{\star}{n}}(\QQ)$ are not jointly typical. \\
\end{itemize}

\textit{Expected error probability.} We introduce the parameter $\varepsilon_1>0$, in order to provide an upper bound on the expected error probability. For all $\varepsilon_1>0$ there exists an $\bar{n} \in \N$ such that for all $n\geq\bar{n}$, the expected probability of the following error events are bounded by $\varepsilon_1$:
\begin{eqnarray}
&&\E_c\bigg[ \PP\bigg( \forall  m \in \mc{M}  , 
\big(U^n, W^n(m) \big) \notin A_{\varepsilon}^{{\star}{n}}(\QQ) \bigg)\bigg]  \leq \varepsilon_1, \nonumber \\&&\label{eq:AchievProba1} \\
&&\E_c\bigg[ \PP\bigg(  \exists m'\neq  m  ,
\big(Y^n , X^n(m') \big) \in A_{\varepsilon}^{{\star}{n}}( \QQ)\bigg)\bigg]   \leq \varepsilon_1, \nonumber \\&&\label{eq:AchievProba2}\\
&&\E_c\bigg[ \PP\bigg(\big(U^n ,  W^n(m) , V^n \big)    \notin A_{\varepsilon}^{{\star}{n}} (\QQ) \bigg)\bigg]  \leq \varepsilon_1.  \label{eq:AchievProba3} 
\end{eqnarray}
\eqref{eq:AchievProba1} comes from \eqref{eq:AchievabilitySEC1} and \cite[pp. 208, Covering Lemma]{ElGammalKim(book)11}.\\
\eqref{eq:AchievProba2} comes from \eqref{eq:AchievabilitySEC2} and \cite[pp. 46, Packing Lemma]{ElGammalKim(book)11}.\\
\eqref{eq:AchievProba3} comes from the properties of typical sequences, stated in \cite[pp. 27]{ElGammalKim(book)11}, and from \eqref{eq:AchievProba1} and \eqref{eq:AchievProba2}.\\

This proves that for all $\varepsilon_1>0$, there exists a $\bar{n}\in \N$ such that for all $n\geq\bar{n}$,  there exists a code  $c^{\star} = (f^{\star},g^{\star})\in \mc{C}(n)$ such that:
\begin{eqnarray}
&& \PP_{c^{\star}} \bigg(\Big|\Big|Q^{n }(u,w,v) - \QQ(u,w,v) \Big|\Big|_{1}\geq \varepsilon \bigg) \nonumber \\
&=& \PP_{c^{\star}}\bigg(\big(U^n ,  W^n(m) ,V^n \big)    \notin A_{\varepsilon}^{{\star}{n}} (\QQ) \bigg) \leq \varepsilon_1.  \label{eq:AchievProba6} 
\end{eqnarray}
We denote by $(\sigma^{\star},\tau^{\star})$, the strategies of $P_1$ and $P_2$ corresponding to the coding scheme $c^{\star} = (f^{\star},g^{\star})\in \mc{C}(n)$ and we denote by $ \bar{\phi}_1 = \max_{u,v} \Big| \phi_1(u,v) \Big|$, the maximal utility of $P_1$.
\begin{eqnarray*}
&&\bigg| \Phi_1^n(\sigma^{\star},\tau^{\star}) - \Phi_1^{\star} \bigg| \nonumber \\
&=&  \Bigg|\sum_{u,v}     \phi_1(u,v) \times \bigg( \E  \Big[ Q^n(u,v) \Big]  -  \QQ(u,v)  \bigg) \Bigg|\\
&\leq&  \bar{\phi}_1 \cdot  \sum_{u,v}    \bigg| \E  \Big[ Q^n(u,v) \Big]  -  \QQ(u,v)  \bigg|
\leq  \bar{\phi}_1 \cdot \varepsilon.
\end{eqnarray*} 
Hence the pair of utilities $\big(\Phi_1^n(\sigma^{\star},\tau^{\star}) , \Phi_2^n(\sigma^{\star},\tau^{\star}) \big)$ converges to the utilities $\big( \Phi_1^{\star} , \Phi_2^{\star} \big) $, as $n \longrightarrow + \infty$.


\subsection{Equality in the Information Constraint}\label{sec:EqualityIC}

We consider a target distribution $ \PP_{\sf{u}}(u) \times \QQ(w | u) \times \QQ(v |w)$ with equality in the information constraint:
\begin{eqnarray}
 \max_{\PP(x)} I( X; Y )  -   I( W ;U  ) = 0.
\end{eqnarray}
\bd{First case: the channel capacity is strictly positive.}
\begin{eqnarray}
\max_{\PP(x)} I( X; Y )  > 0.
\end{eqnarray}
We consider an auxiliary distribution $\widetilde{\QQ}(w|u)=\widetilde{\QQ}(w)$ such that $W$ is independent of $U$ and we denote by $I_{\widetilde{\QQ}}( W ;U  )=0$, the corresponding mutual information. The information constraint for $ \PP_{\sf{u}}(u) \times \widetilde{\QQ}(w|u)$ is strictly positive:
\begin{eqnarray}
\max_{\PP(x)} I( X; Y )  -   I_{\widetilde{\QQ}}( W ;U  )  &=&\max_{\PP(x)} I( X; Y ) >0. \label{eq:ZeroIC1} 
\end{eqnarray}
We construct a sequence  $\big\{\QQ^k(u,w,v)\big\}_{k\in\N^*}$ of convex combination between $\QQ(u,w,v)$ and $ \PP_{\sf{u}}(u) \times  \widetilde{\QQ}(w|u) \times \QQ(v |w)$:
\begin{eqnarray}
\QQ^k(u,w,v) &=& \frac{1}{k} \bigg( (k-1) \cdot \QQ(u,w,v) \nonumber \\
&+&  \PP_{\sf{u}}(u) \times \widetilde{\QQ}(w)  \times \QQ(v|w) \bigg).
\end{eqnarray}
The information constraint corresponding to $\QQ^k(u,w,v)$ is strictly positive, for all $k \in\N^*$:
\begin{eqnarray}
&&\max_{\PP(x)} I( X; Y ) -   I_{\QQ^k}( W ;U  )  \nonumber \\
&\geq& \frac{1}{n} \cdot\bigg( (n-1) \cdot \big( \max_{\PP(x)} I( X; Y ) -   I_{\QQ}( W ;U  ) \big)\nonumber \\
&+&   \big( \max_{\PP(x)} I( X; Y ) -   I_{\widetilde{\QQ}}( W ;U  ) \big) \bigg)\\
&\geq& \frac{1}{n} \cdot   \big( \max_{\PP(x)} I( X; Y ) -   I_{\widetilde{\QQ}}( W ;U  ) \big)  >0.
\end{eqnarray}
Then, for all $k \in\N^*$ the distribution $\QQ^k(u,w,v)$ is achievable by using the coding scheme stated in Sec.\ref{sec:ExistenceOptimalStrategy} and converges to the target distribution $\QQ(u,w,v)$, as $k$ goes to $+\infty$. This proves that the limit distribution $\QQ(u,w,v) $ is achievable.

\bd{Second case: the channel capacity is equal to zero.} 
This implies that the  random variables $U$ and $W$ of the target distribution $ \PP_{\sf{u}}(u) \times \QQ(w | u)\times  \QQ(v |w)$ are independent, hence the target distribution $\QQ(u,w,v)$ decomposes like:
\begin{eqnarray} 
\PP_{\sf{u}}(u) \times \QQ(w | u)  \times \QQ(v |w) =\PP_{\sf{u}}(u) \times \QQ(w )  \times \QQ(v |w) . \label{eq:ZeroIC5}
 \end{eqnarray}
This target distribution   \eqref{eq:ZeroIC5} is achievable by the decoder generating $(W^n,V^n)$ with the  i.i.d. distribution $ \QQ(w )  \times \QQ(v |w)$.


\subsection{Unilateral Deviation of the Second Player}\label{sec:UnilateralDev2}

The target joint distribution $\QQ(u,w,v) =  \PP_{\sf{u}}(u) \times \QQ(w | u)\times   \QQ(v |w)$ of Definition \ref{def:Q}, satisfies the information constraint of $\Q_0$ and the strategic compatibility condition of $\Q_2$. We consider the coding and decoding functions $c^{\star}=(f^{\star},g^{\star})$ presented in App. \ref{sec:ExistenceOptimalStrategy}, for distribution $\QQ(u,w,v)$. In this section, we prove that the decoding function $g^{\star}$ combined with a symbol-best-reply $\QQ(v|w)$ is $\varepsilon-$optimal for $P_2$.
\begin{figure*}[ht!]
\begin{small}
\begin{eqnarray}
&&\max_{ \PP(v^n|y^n, c^{\star},E=0) }  \sum_{u^n,x^n,y^n,\atop v^n,w^n} \PP(u^n,x^n,y^n,w^n| c^{\star},E=0)   \times \PP(v^n| y^n, c^{\star},E=0)  \cdot \frac{1}{n} \sum_{i=1}^n \phi_2(u_i,v_i)\nonumber \\
&\leq&\max_{ \PP(v^n|w^n,y^n, c^{\star},E=0) }  \sum_{u^n,x^n,y^n,\atop v^n,w^n} \PP(u^n,x^n,y^n,w^n| c^{\star},E=0) 
\times\PP(v^n| w^n,y^n, c^{\star},E=0)  \cdot \frac{1}{n} \sum_{i=1}^n \phi_2(u_i,v_i) \label{eq:MaxUtility2}  \\
&=&\max_{ \PP(v^n|w^n,y^n, c^{\star},E=0) }  \sum_{u^n,x^n,y^n,\atop v^n,w^n} \PP(x^n,y^n,w^n| c^{\star},E=0) \times \PP(u^n|w^n, c^{\star},E=0)  \times\PP(v^n| w^n,y^n, c^{\star},E=0)  \cdot \frac{1}{n} \sum_{i=1}^n \phi_2(u_i,v_i) \nonumber \\&& \label{eq:MaxUtility3}  \\
&=&  \sum_{w^n} \PP(w^n| c^{\star},E=0) \times\max_{ \PP(v^n|w^n,c^{\star},E=0) } \sum_{v^n}  \PP(v^n| w^n,c^{\star},E=0)  \times  \sum_{u^n} \PP(u^n|w^n, c^{\star},E=0)  \cdot \frac{1}{n} \sum_{i=1}^n \phi_2(u_i,v_i)\label{eq:MaxUtility5} \\
&\leq&  \sum_{w^n} \PP(w^n| c^{\star},E=0) \times\max_{ \PP(v^n|w^n,c^{\star},E=0) } \sum_{v^n}  \PP(v^n| w^n,c^{\star},E=0)  \times  \sum_{u^n} \bigg( \prod_{i=1}^n Q(u_i | w_i) \bigg)  \cdot \frac{1}{n} \sum_{i=1}^n \phi_2(u_i,v_i) + 2\sqrt{\ln2\varepsilon} \cdot \bar{\phi}_2 \label{eq:MaxUtility6} \\
&=&  \sum_{w^n} \PP(w^n| c^{\star},E=0)\sum_{w} Q^n(w)  \max_{\PP(v|w)}  \sum_{v} \PP(v|w)    \sum_{u\in \mc{U}  }\QQ(u|w) \cdot   \phi_2(u,v) + 2\sqrt{\ln2\varepsilon} \cdot \bar{\phi}_2 \label{eq:MaxUtility7} \\
&\leq& \sum_{w} \QQ(w)  \max_{\PP(v|w)}  \sum_{v} \PP(v|w)    \sum_{u\in \mc{U}  }\QQ(u|w) \cdot   \phi_2(u,v) + (2\sqrt{\ln2\varepsilon}+  \varepsilon) \cdot \bar{\phi}_2 \label{eq:MaxUtility8} \\
&=& \sum_{u,w,v} \QQ(u,w)\cdot  \QQ(v|w)  \cdot   \phi_2(u,v)  + (2\sqrt{\ln2\varepsilon}+  \varepsilon) \cdot \bar{\phi}_2. \label{eq:MaxUtility9}
\end{eqnarray}
\end{small}
\end{figure*}
We introduce the random  event of error $E \in \{0,1\}$ defined by:
\begin{eqnarray*}
E = \Bigg\{
\begin{array}{lll}
0 &\text{ if }&  (U^n,W^n,V^n) \in A_{\varepsilon}^{{\star}{n}}(\QQ)\;\; \text{ and }  \hat{M} = M,\\
1 &\text{ if }&(U^n,W^n,V^n) \notin A_{\varepsilon}^{{\star}{n}}(\QQ) \;\; \text{ or } \;\; \hat{M} \neq M.
\end{array}
\Bigg.
\end{eqnarray*}

The pair of strategies $c^{\star} = (f^{\star},g^{\star})\in \mc{C}(n)$, stated in App. \ref{sec:ExistenceOptimalStrategy}, induces a small error probability $\PP(E=1|c^{\star})\leq \varepsilon$. The expected utility of $P_2$ is upper bounded by:
\begin{eqnarray}
&&  \sum_{u^n,x^n,y^n,v^n\atop w^n,E} \PP(u^n,x^n,y^n,w^n,v^n,E| c^{\star})  \cdot \frac{1}{n} \sum_{i=1}^n \phi_2(u_i,v_i)  \nonumber \\
&=& \PP(E=0| c^{\star})   \sum_{u^n,x^n,y^n,\atop v^n,w^n} \PP(u^n,x^n,y^n,w^n,v^n| c^{\star},E=0) \nonumber \\
&&  \times \; \frac{1}{n} \sum_{i=1}^n \phi_2(u_i,v_i) \nonumber \\
&+&  \PP(E=1| c^{\star})   \sum_{u^n,x^n,y^n,\atop v^n,w^n} \PP(u^n,x^n,y^n,w^n,v^n| c^{\star},E=1)   \nonumber \\
&&  \times \; \frac{1}{n} \sum_{i=1}^n \phi_2(u_i,v_i)  \label{eq:StrategyCode4} \\
&\leq&\sum_{u^n,x^n,y^n,\atop v^n,w^n} \PP(u^n,x^n,y^n,w^n,v^n| c^{\star},E=0)   \nonumber \\
&&  \times \; \frac{1}{n} \sum_{i=1}^n \phi_2(u_i,v_i) + \PP(E=1| c^{\star})  \times \bar{ \phi}_2  . \label{eq:StrategyCode5}
\end{eqnarray}
We denote by $\bar{\phi}_2 = \max_{u,v}\Big|\phi_2(u,v)\Big|$, the maximal utility of $P_2$. In the following, we assume that $P_2$ chooses the optimal sequence $V^n$ based on her observation $Y^n$, on the knowledge of the code $c^{\star}$, on the hypothesis that there is no errors $E=0$ and on the decoded sequence $W^n$.  We prove that the decoding function $g^{\star}$ presented in App. \ref{sec:ExistenceOptimalStrategy}, is an $\varepsilon-$best-reply for $P_2$. 

Equation \eqref{eq:MaxUtility2} comes from the hypothesis  $E=0$ of correct decoding of the sequence $W^n$. The decoder maximizes over $\PP(v^n| w^n,y^n, c^{\star},E=0)$ instead of $\PP(v^n| y^n, c^{\star},E=0)$.\\
Equation \eqref{eq:MaxUtility3} comes from the  Markov chain $U^n -\!\!\!\!\minuso\!\!\!\!- W^n -\!\!\!\!\minuso\!\!\!\!- X^n -\!\!\!\!\minuso\!\!\!\!-Y^n$ of the coding process $c^{\star}$ stated in App. \ref{sec:ExistenceOptimalStrategy}, that induces the following equality $ \PP(u^n| w^n,x^n,y^n, c^{\star},E=0) = \PP(u^n| w^n , c^{\star},E=0) $.\\
Equation \eqref{eq:MaxUtility5} comes from taking the sum over $(x^n, y^n)$ and removing the sequence $y^n$ from  $\PP(v^n| w^n,c^{\star},E=0)$, since $y^n$ is not involved in the criteria $\sum_{i=1}^n \phi_2(u_i,v_i) $.\\
Equation \eqref{eq:MaxUtility6} is due to Lemmas \ref{lemma1} and \ref{lemma:PosteriorBeliefs}, stated in App. \ref{sec:Lemmas}.\\
Equation \eqref{eq:MaxUtility7} comes from Lemma \ref{lemma2}, stated in App. \ref{sec:Lemmas}.\\
Equation \eqref{eq:MaxUtility8} comes from the hypothesis $E=0$ that implies the empirical distribution $Q^n(w)$ of sequences $w^n$ is close to the target $\QQ(w)$.\\
Equation \eqref{eq:MaxUtility9} comes from the definition of the set $\Q_2$ that requires the distribution $\QQ(v|w)$ maximizes $\max_{\PP(v|w)}  \sum_{v} \PP(v|w)    \sum_{u\in \mc{U}  }\QQ(u|w) \cdot   \phi_2(u,v)$.

We proved that the equilibrium condition is satisfied:
\begin{eqnarray}
 \Phi_2^n\big(\sigma^{\star},\tau^{\star}(\sigma^{\star})\big) &\geq&\max_{\tilde{\tau}}\; \Phi_2^n\big(\sigma^{\star},\tilde{\tau}\big)  -  \varepsilon .
\end{eqnarray}


\subsection{Lemmas}\label{sec:Lemmas}

To simplify the notations of Lemma \ref{lemma1} and \ref{lemma2}, we remove the conditioning over the code $c^{\star}$ and the event $E=0$ in the probabilities. 
\begin{lemma}\label{lemma1}
The following expression satisfies: 
\begin{small}
\begin{eqnarray}
&& \Bigg| \sum_{w^n,v^n} \PP(w^n,v^n)  \sum_{u^n}\PP(u^n|w^n) \cdot \frac{1}{n} \sum_{i=1}^n \phi_2(u_i,v_i) \nonumber\\
&-&  \sum_{w^n,v^n} \PP(w^n,v^n)  \sum_{u^n} \bigg( \prod_{i=1}^n Q(u_i|w_i) \bigg)  \cdot \frac{1}{n} \sum_{i=1}^n \phi_2(u_i,v_i) \Bigg| \nonumber\\
& \leq& \bar{\phi}_2 \cdot  2\sqrt{\ln2\varepsilon}.   \label{eq:theoMaximization5} 
\end{eqnarray}
\end{small}
\end{lemma}

\begin{lemma}\label{lemma2}
For all sequence $w^n\in \mc{W}^n$, we have this equality: 
\begin{small}
\begin{eqnarray*}
 \max_{\PP(v^n|w^n)} \sum_{v^n} \PP(v^n | w^n)   \sum_{u^n  } \bigg( \prod_{i=1}^n Q(u_i | w_i) \bigg) \cdot \frac{1}{n} \sum_{i=1}^n \phi_2(u_i,v_i) \nonumber \\
=  \sum_{w} Q^n(w)  \max_{\PP(v|w)}  \sum_{v} \PP(v|w)    \sum_{  u\in \mc{U} }Q(u|w) \cdot   \phi_2(u,v)   \label{eq:theoMaximizationEval5}.
\end{eqnarray*}
\end{small}
\end{lemma}
The proof of Lemma \ref{lemma1} is based on Lemma \ref{lemma:PosteriorBeliefs} and the proof of Lemma \ref{lemma2} comes from the hypothesis $E=0$, of jointly typical sequences $(u^n,w^n,v^n) \in A_{\varepsilon}^{{\star}{n}}(\QQ)$.

\begin{lemma}[Posteriors beliefs]\label{lemma:PosteriorBeliefs}
The coding scheme  $c^{\star} = (f^{\star},g^{\star})\in \mc{C}(n)$ described in App. \ref{sec:ExistenceOptimalStrategy} satisfies:
\begin{eqnarray}
&&\E_{W^n} \Bigg[  \frac{1}{n}   \cdot \sum_{i=1}^n  \Big|\Big| \PP(U_i|W^n,c^{\star},E=0) - \QQ(U_i|W_i)\Big|\Big|_1 \Bigg] \nonumber \\&&\leq 2\sqrt{\ln2\varepsilon}. \label{eq:LemmaPosteriorBeliefs}
\end{eqnarray}
\end{lemma}
Lemma \ref{lemma:PosteriorBeliefs} corresponds to the notion of ``Strategic Distance'' introduced in \cite{GossnerVieille02} and in the proof of \cite[Lemma 36]{GossnerTomala06} that implies the main result of \cite{GossnerTomala07} and \cite{GossnerLarakiTomala09}. 

\begin{proof}[Lemma \ref{lemma:PosteriorBeliefs}] We consider the code $c^{\star} = (f^{\star},g^{\star})\in \mc{C}(n)$, stated in App. \ref{sec:ExistenceOptimalStrategy} and we assume that the sequences $(U^n,W^n,V^n) \in A_{\varepsilon}^{{\star}{n}}(\QQ)$ are jointly typical, \text{i.e.} the error event is $E=0$. We provide an upper bound on the $L_1$ distance based on Pinsker's and Jensen's inequalities. We denote by $D(\PP||\QQ)$ the K-L divergence between distributions $\PP$ and $\QQ$.
\begin{footnotesize}
\begin{eqnarray}
&&\E_{W^n} \Bigg[  \frac{1}{n}   \cdot \sum_{i=1}^n  \Big|\Big| \PP(U_i|W^n,c^{\star},E=0) - \QQ(U_i|W_i)\Big|\Big|_1 \Bigg] \nonumber\\
&=& \sum_{w^n} \PP(w^n | c^{\star},E=0 )\nonumber\\
&\times& \frac{1}{n}   \sum_{i=1}^n \Big|\Big| \PP(U_i|w^n,c^{\star},E=0) - \QQ(U_i|w_i)  \Big|\Big|_1 \label{eq:L1distance1} \\
&\leq& \sum_{w^n} \PP(w^n | c^{\star},E=0 ) \nonumber\\
&\times&  \frac{1}{n}   \sum_{i=1}^n \sqrt{2 \ln 2 \cdot D\bigg(  \PP(U_i|w^n,c^{\star},E=0) \bigg| \bigg|   \QQ(U_i|w_i) \bigg) }\label{eq:L1distance3} \\
&\leq&  \sqrt{2 \ln 2\sum_{w^n} \PP(w^n | c^{\star},E=0 )} \nonumber\\
&\times& \sqrt{ \frac{1}{n}   \sum_{i=1}^n  D\bigg(  \PP(U_i|w^n,c^{\star},E=0) \bigg| \bigg|   \QQ(U_i|w_i) \bigg) } \label{eq:L1distance4}\\
&\leq&  \sqrt{2 \ln 2 \cdot \frac{1}{n}  \sum_{i=1}^n  \E_{W^n} \Bigg[ D\bigg(  \PP(U_i|W^n,c^{\star},E=0) \bigg| \bigg|   \QQ(U_i|W_i) \bigg)\Bigg] }. \nonumber \\&&\label{eq:L1distance5}
\end{eqnarray}
\end{footnotesize}
In equation \eqref{eq:L1distance1}, the $L_1$ distance regards $U_i$.\\
Equation \eqref{eq:L1distance3} comes from Pinsker's inequality, \cite[pp. 370]{cover-book-2006}.\\
Equation \eqref{eq:L1distance4} comes from Jensen's inequality for $x\mapsto\sqrt{x}$.\\

\begin{eqnarray}
&&\frac{1}{n}  \sum_{i=1}^n  \E_{W^n} \Bigg[ D\bigg(  \PP(U_i|W^n,c^{\star},E=0) \bigg| \bigg|   \QQ(U_i|W_i) \bigg)\Bigg] \nonumber \\
&=&  \frac{1}{n}     \sum_{(u^n,w^n)\in A_{\varepsilon}^{{\star}{n}}(\QQ)} \PP(u^n,w^n | c^{\star},E=0 ) \nonumber \\
&\times&  \log_2 \frac{1}{\prod_{i=1}^n  \QQ(u_i|w_i)  }-\frac{1}{n}  \sum_{i=1}^n  H(U_i|W^n,c^{\star},E=0)\label{eq:KLdistance2}  \nonumber \\&&\\
&\leq&  \frac{1}{n}     \sum_{(u^n,w^n)\in A_{\varepsilon}^{{\star}{n}}(\QQ)} \PP(u^n,w^n | c^{\star},E=0 )\nonumber \\
&\times&  n \cdot \bigg( H(U |  W) + \varepsilon \bigg) -  \frac{1}{n}   H(U^n|W^n,c^{\star},E=0)\label{eq:KLdistance3} \\
&=&  \frac{1}{n}   I(U^n;W^n|c^{\star},E=0) -  I(U ;  W)  + \varepsilon \label{eq:KLdistance4}\\
&\leq& \log|\mc{M}| -  I(U ;  W)  + \varepsilon\label{eq:KLdistance5} \\
&\leq& I(U ;  W)  + \varepsilon  -  I(U ;  W)  + \varepsilon\label{eq:KLdistance6} \\
&\leq& 2 \varepsilon. \label{eq:KLdistance7}
\end{eqnarray}
Equation \eqref{eq:KLdistance2} is the definition of the K-L divergence.\\
Equation \eqref{eq:KLdistance3} comes from the property of typical sequences $(U^n,W^n)\in A_{\varepsilon}^{{\star}{n}}(\QQ)$ in \cite[pp. 26]{ElGammalKim(book)11} and of the  entropy.\\
Equation \eqref{eq:KLdistance4} comes from the i.i.d. property of the source $U$.\\
Equations \eqref{eq:KLdistance5} and \eqref{eq:KLdistance6} come from the cardinality of the codebook $|\mc{M}|$, introduced in App. \ref{sec:ExistenceOptimalStrategy}.\\

Equation \eqref{eq:KLdistance4} involves the information leakage $ \frac{1}{n}   I(U^n;W^n|c^{\star},E=0)$ corresponding to the amount of information received by $P_2$, regarding the source $U^n$. The information leakage induced by the coding scheme for empirical coordination, is investigated in \cite{LeTreustBloch(ISIT)16}. \end{proof}

\bibliographystyle{ieeetr}

\begin{thebibliography}{}

\end{thebibliography}


\begin{thebibliography}{40}

\bibitem{GoHerNey06}
O.~Gossner, P.~Hernandez, and A.~Neyman, ``Optimal use of communication
  resources,'' {\em Econometrica}, vol.~74, pp.~1603--1636, Nov. 2006.

\bibitem{GossnerTomala07}
O.~Gossner and T.~Tomala, ``Secret correlation in repeated games with imperfect
  monitoring,'' {\em Mathematics of Operation Research}, vol.~32, no.~2,
  pp.~413--424, 2007.

\bibitem{GossnerTomala06}
O.~Gossner and T.~Tomala, ``Empirical distributions of beliefs under imperfect
  observation,'' {\em Mathematics of Operation Research}, vol.~31, no.~1,
  pp.~13--30, 2006.

\bibitem{GossnerLarakiTomala09}
O.~Gossner, R.~Laraki, and T.~Tomala, ``Informationally optimal correlation,''
  {\em Mathematical Programming}, vol.~116, no.~1-2, pp.~147--172, 2009.

\bibitem{GossnerVieille02}
O.~Gossner and N.~Vieille, ``How to play with a biased coin?,'' {\em Games and
  Economic Behavior}, vol.~41, no.~2, pp.~206--226, 2002.

\bibitem{KramerSavari07}
G.~Kramer and S.~Savari, ``Communicating probability distributions,'' {\em
  Information Theory, IEEE Transactions on}, vol.~53, no.~2, pp.~518 -- 525,
  2007.

\bibitem{CuffPermuterCover10}
P.~Cuff, H.~Permuter, and T.~Cover, ``Coordination capacity,'' {\em Information
  Theory, IEEE Transactions on}, vol.~56, no.~9, pp.~4181--4206, 2010.

\bibitem{Cuff(ImplicitCoordination)11}
P.~Cuff and L.~Zhao, ``Coordination using implicit communication,'' {\em
  Information Theory Workshop (ITW), IEEE}, pp.~467-- 471, 2011.

\bibitem{CuffSchieler11}
P.~Cuff and C.~Schieler, ``Hybrid codes needed for coordination over the
  point-to-point channel,'' in {\em Communication, Control, and Computing
  (Allerton), 2011 49th Annual Allerton Conference on}, pp.~235--239, Sept
  2011.

\bibitem{LetreustZaidiLasaulce(Allerton)11}
M.~Le~Treust, A.~Zaidi, and S.~Lasaulce, ``An achievable rate region for the
  broadcast wiretap channel with asymmetric side information,'' {\em IEEE Proc.
  of the 49th Allerton conference, Monticello, Illinois}, pp.~68 -- 75.

\bibitem{LeTreust(EmpiricalCoordination)14}
M.~Le~Treust, ``Empirical coordination for the joint source-channel coding
  problem,'' {\em submitted to IEEE Trans. on Information Theory,
  http://arxiv.org/abs/1406.4077}, 2014.

\bibitem{LarrousseLasaulceBloch(IT)14}
B.~Larrousse, S.~Lasaulce, and M.~Bloch, ``Coordination in distributed networks
  via coded actions with application to power control,'' {\em Submitted to IEEE
  Transactions on Information Theory, http://arxiv.org/abs/1501.03685}, 2014.

\bibitem{LeTreust(CorrelationITW)14}
M.~Le~Treust, ``Correlation between channel state and information source with
  empirical coordination constraint,'' in {\em IEEE Information Theory Workshop
  (ITW)}, pp.~272--276, Nov 2014.

\bibitem{LeTreust(ISIT-TwoSided)15}
M.~Le~Treust, ``Empirical coordination with two-sided state information and
  correlated source and state,'' in {\em IEEE International Symposium on
  Information Theory (ISIT)}, 2015.

\bibitem{LeTreust(ISITfeedbacks)15}
M.~Le~Treust, ``Empirical coordination with channel feedback and strictly
  causal or causal encoding,'' in {\em IEEE International Symposium on
  Information Theory (ISIT)}, 2015.

\bibitem{LarrousseLasaulceWigger(ITW)15}
B.~Larrousse, S.~Lasaulce, and M.~Wigger, ``Coordinating partially-informed
  agents over state-dependent networks,'' {\em IEEE Information Theory Workshop
  (ITW)}, 2015.

\bibitem{BlascoThobabenSkoglund12}
R.~Blasco-Serrano, R.~Thobaben, and M.~Skoglund, ``Polar codes for coordination
  in cascade networks,'' {\em in Proc. of the International Zurich Seminar on
  Communication, Zurich, Switzerland}, pp.~55 -- 58, March 2012.

\bibitem{ChouBlochKliewer15}
R.~Chou, M.~Bloch, and J.~Kliewer, ``Polar coding for empirical and strong
  coordination via distribution approximation,'' in {\em Information Theory
  Proceedings (ISIT), 2015 IEEE Internat. Symp. on}, June 2015.

\bibitem{ChouBlochKliewer16}
R.~Chou, M.~Bloch, and J.~Kliewer, ``Empirical and strong coordination via soft
  covering with polar codes,'' {\em submitted to IEEE Transactions on
  Information Theory, http://arxiv.org/abs/1608.08474}, 2016.
  
\bibitem{CerviaLuzziBlochLeTreust16}
G.~Cervia, L.~Luzzi, M.~R. Bloch, and M.~L. Treust, ``Polar coding for
  empirical coordination of signals and actions over noisy channels,'' in {\em
  in Proc. IEEE Information Theory Workshop (ITW)}, 2016.

\bibitem{SchielerCuff(RateDistortion14)}
C.~Schieler and P.~Cuff, ``Rate-distortion theory for secrecy systems,'' {\em
  IEEE Trans. on Information Theory}, vol.~60, pp.~7584--7605, Dec 2014.

\bibitem{LeTreustBloch(ISIT)16}
M.~Le~Treust and M.~Bloch, ``Empirical coordination, state masking and state
  amplification: Core of the decoder's knowledge,'' {\em Proceedings of the
  IEEE International Symposium on Information Theory (ISIT)}, 2016.

\bibitem{Nash51}
J.~Nash, ``Non-cooperative games,'' {\em Annals of Mathematics}, vol.~54,
  pp.~286--295, 1951.

\bibitem{stackelberg-book-1934}
H.~von Stackelberg, {\em Marketform und Gleichgewicht}.
\newblock Oxford University Press, 1934.

\bibitem{CrawfordSobel1982StrategicInformation}
V.~P. Crawford and J.~Sobel, ``{Strategic Information Transmission},'' {\em
  Econometrica}, vol.~50, no.~6, pp.~1431--1451, 1982.

\bibitem{Forges94}
F.~Forges, ``Non-zero-sum repeated games and information transmission,'' {\em
  in: N. Meggido, Essays in Game Theory in Honor of Michael Maschler,
  Springer-Verlag}, no.~6, pp.~65--95, 1994.

\bibitem{KamenicaGentzkow11}
E.~Kamenica and M.~Gentzkow, ``Bayesian persuasion,'' {\em American Economic
  Review}, vol.~101, pp.~2590 -- 2615, 2011.

\bibitem{AM95}
R.~Aumann and M.~Maschler, {\em Repeated Games with Incomplete Information}.
\newblock MIT Press, Cambrige, MA, 1995.

\bibitem{sorin-book-2002}
S.~Sorin, {\em A First Course on Zero-Sum Repeated Games}, vol.~37 of {\em
  Math\'{e}matiques et Applications}.
\newblock Springer, 2002.

\bibitem{MertensSorinZamir15}
J.~Mertens, S.~Sorin, and S.~Zamir, {\em Repeated Games}.
\newblock Cambridge University Press, 2015.

\bibitem{Laraki01}
R.~Laraki, ``The splitting game and applications,'' {\em International Journal
  of Game Theory}, vol.~30, pp.~359--376, 2001.

\bibitem{RenaultSolanVieille16}
J.~Renault, E.~Solan, and N.~Vieille, ``Optimal dynamic information
  provision,'' {\em
  http://www.lse.ac.uk/statistics/events/2015-16-Seminar-Series/Optimal-Dynamic-Information-Provision.pdf},
  February 2016.

\bibitem{ely2015beeps}
J.~Ely, ``Beeps,'' {\em Manuscript, Department of Economics, Northwestern
  University}, 2015.

\bibitem{JACKSONSONNENSCHEIN07}
M.~O. Jackson and H.~F. Sonnenschein, ``Overcoming incentive constraints by
  linking decisions,'' {\em Econometrica}, vol.~75, pp.~241 -- 257, January
  2007.

\bibitem{AkyolLangbortBasar15}
E.~Akyol, C.~Langbort, and T.~Ba\c{s}ar, ``Strategic compression and transmission
  of information,'' in {\em Information Theory Workshop - Fall (ITW), 2015
  IEEE}, pp.~219--223, Oct 2015.

\bibitem{AkyolLangbortBasar16}
E.~Akyol, C.~Langbort, and T.~Ba\c{s}ar, ``On the role of side information in
  strategic communication,'' in {\em 2016 IEEE International Symposium on
  Information Theory (ISIT)}, pp.~1626--1630, July 2016.

\bibitem{SaritasYükselGezici16}
S.~Sar{\i}ta\c{s}, S.~Yüksel, and S.~Gezici, ``Dynamic signaling games under nash and
  stackelberg equilibria,'' in {\em 2016 IEEE International Symposium on
  Information Theory (ISIT)}, pp.~1631--1635, July 2016.
  
  \bibitem{Belmega-TWC-2009}
E.~V. Belmega, S.~Lasaulce, and M.~Debbah, ``Power allocation games for mimo
  multiple access channels with coordination,'' {\em IEEE Trans. on Wireless
  Communications}, vol.~8, no.~6, pp.~3182--3192, 2009.
  

\bibitem{ElGammalKim(book)11}
A.~E. Gamal and Y.-H. Kim, {\em Network Information Theory}.
\newblock Cambridge University Press, Dec. 2011.

\bibitem{cover-book-2006}
T.~M. Cover and J.~A. Thomas, {\em Elements of information theory}.
\newblock New York: 2nd. Ed., Wiley-Interscience, 2006.




\end{thebibliography}
%

%

\end{document}